\documentclass[acmsmall]{acmart}

\acmDOI{10.1145/3660807}
\acmYear{2024}
\copyrightyear{2024}
\acmSubmissionID{fse24main-p823-p}
\acmJournal{PACMSE}
\acmVolume{1}
\acmNumber{FSE}
\acmArticle{100}
\acmMonth{7}
\received{2023-09-28}
\received[accepted]{2024-04-16}

\AtBeginDocument{
  \providecommand\BibTeX{{
    \normalfont B\kern-0.5em{\scshape i\kern-0.25em b}\kern-0.8em\TeX}}}

\usepackage{xparse}
\usepackage{xspace}
\usepackage{comment}
\usepackage{xcolor, soul}
\usepackage{xcolor,colortbl}
\usepackage{lipsum}% http://ctan.org/pkg/lipsum
\usepackage{tcolorbox}
\usepackage{enumitem, kantlipsum}
\usepackage{filecontents,lipsum}
\usepackage[normalem]{ulem} % For strikethrough, normalem option to keep \emph{} as italic

\newcommand{\original}[1]{}
\setlist[itemize]{leftmargin=*}
\usepackage{graphicx}
\usepackage{hyperref}
\usepackage{mdframed}
\usepackage{amsmath}
\usepackage{wrapfig}

\newcommand{\sm}{San Martino's token overlapping metrics}
\newcommand{\ka}{Krippendorff’s alpha}

\usepackage{multirow}

\newcommand{\mr}[1]{#1}

\usepackage{lipsum}
\usepackage{xcolor}
\usepackage{caption}
\usepackage{tabularray}
\UseTblrLibrary{booktabs}
\usepackage{makecell}
\usepackage{booktabs}
\usepackage{siunitx}

\newcommand{\yellow}[1]{\colorbox{tab_highlight}{#1}}

\definecolor{dr}{HTML}{8b0000}

\newcommand{\modelkeyword}[2]{\colorbox{lightblue}{#2}}

\newcommand{\bothkeyword}[2]{\colorbox{lightblue}{#2}}

\usepackage{bbm}

\usepackage{svg}
\usepackage{booktabs}
\usepackage{array}
\usepackage{diagbox}
\usepackage{titlesec}
\newcommand{\PreserveBackslash}[1]{\let\temp=\\#1\let\\=\temp}
\newcolumntype{C}[1]{>{\PreserveBackslash\centering}p{#1}}
\newcolumntype{R}[1]{>{\PreserveBackslash\raggedleft}p{#1}}
\newcolumntype{L}[1]{>{\PreserveBackslash\raggedright}p{#1}}

\usepackage{MnSymbol}
\definecolor{Gray}{gray}{0.9}
\definecolor{light_green}{HTML}{c4edc9}
\definecolor{light_red}{HTML}{fddad9}
\definecolor{light_blue}{HTML}{cad7fb}
\definecolor{light_yellow}{HTML}{FFE599}
\definecolor{lightyellow}{HTML}{ffffed}
\definecolor{lightpurple}{HTML}{E6E6FA}

\DeclareRobustCommand{\hlgreen}[1]{{\sethlcolor{light_green}\hl{#1}}}
\DeclareRobustCommand{\hlblue}[1]{{\sethlcolor{light_blue}\hl{#1}}}

\definecolor{tab_highlight}{rgb}{0.99,0.80,0.49}

\newcolumntype{N}[1]{>{\centering\arraybackslash}m{#1cm}}

\usepackage{color, colortbl}

\newcommand{\website}{https://github.com/BonanKou/Attention-Alignment-Empirical-Study}

\newcounter{finding}
\newenvironment{finding}
{
    \refstepcounter{finding}
	\begin{mdframed}[
    	nobreak=true,
    	linecolor=black,
    	roundcorner=12pt,
    	backgroundcolor=gray!05,
    	linewidth=0.3pt,
    	leftmargin=0.5em,
    	rightmargin=0.5em,
    	topline=true,
    	bottomline=true,
    	frametitlerule=true,
    	frametitlebackgroundcolor=gray!30,
    	frametitlerulecolor=gray,
    	frametitle=Finding \arabic{finding},
    	frametitleaboveskip=0.3em,
    	frametitlebelowskip=0.35em,
    	skipabove=12pt
	]
}
{
    \end{mdframed}
    \vspace{1em}
}

\usepackage{tcolorbox}

\usepackage{subcaption}

\usepackage{wrapfig}

\usepackage{listings}
\usepackage{xcolor}

\definecolor{codegreen}{rgb}{0,0.6,0}
\definecolor{codegray}{rgb}{0.5,0.5,0.5}
\definecolor{codepurple}{rgb}{0.58,0,0.82}
\definecolor{backcolour}{rgb}{0.95,0.95,0.92}

\definecolor{darkBlue}{rgb}{0.000000,0.000000,0.545098}
\definecolor{darkGreen}{rgb}{0.000000,0.392157,0.000000}
\definecolor{DarkGray}{gray}{0.4}
\definecolor{javared}{rgb}{0.6,0,0} % for strings
\definecolor{javagreen}{rgb}{0.25,0.5,0.35} % comments
\definecolor{javapurple}{rgb}{0.5,0,0.35} % keywords
\definecolor{javadocblue}{rgb}{0.25,0.35,0.75} % javadoc
\definecolor{lightgray}{gray}{0.8}
\definecolor{lightblue}{rgb}{0.63, 0.79, 0.95}

\definecolor{shadecolor}{RGB}{150,150,150}
\definecolor{blueA}{RGB}{204,229,255}
\definecolor{redA}{RGB}{112,0, 0}
\definecolor{RED}{RGB}{255,0, 0}

\lstdefinestyle{mystyle}{
    frame=none,
  xleftmargin=15pt,
  stepnumber=1,
  numbers=left,
  numbersep=5pt,
  stepnumber=1,
  numberstyle=\tiny\bf,%\color[gray]{0.777},
  belowcaptionskip=\bigskipamount,
  captionpos=b,
  escapeinside={*‘}{’*},
  tabsize=5,
  emphstyle={\bf},
  basicstyle=\scriptsize\ttfamily,
  keywordstyle=\color{javapurple}\bfseries,
  stringstyle=\color{javared},
  commentstyle=\color{javagreen},
  morecomment=[s][\color{javadocblue}]{/**}{*/},
  showspaces=false,
  columns=flexible,
  showstringspaces=false,
  morecomment=[l]{//},
  tabsize=2,
  breaklines=true
}

\begin{document}

\title{Do Large Language Models Pay Similar Attention Like Human Programmers When Generating Code?}
% \title{Do Code Generation Models Think Like Us?}
% \subtitle{A Study of Attention Alignment between Large Language Models and Human Programmers}

\author{Bonan Kou}
\orcid{0000-0003-1407-8522}
\affiliation{%
  \institution{Purdue University}
  \city{West Lafayette}
  \country{USA}
}
\email{koub@purdue.edu}

\author{Shengmai Chen}
\authornote{This work was done when Shengmai Chen was an undergraduate student at Purdue University.}
\orcid{0009-0008-8867-6713}
\affiliation{%
  \institution{Brown University}
  \city{Providence}
  \country{USA}
}
\email{shengmai_chen@brown.edu}

\author{Zhijie Wang}
\orcid{0000-0003-4559-5426}
\affiliation{%
  \institution{University of Alberta}
  \city{Edmonton}
  \country{Canada}
}
\email{zhijie.wang@ualberta.ca}

\author{Lei Ma}
\orcid{0000-0002-8621-2420}
\affiliation{%
  \institution{The University of Tokyo}
  \city{Tokyo}
  \country{Japan}
}
\affiliation{%
  \institution{University of Alberta}
  \city{Edmonton}
  \country{Canada}
}
\email{ma.lei@acm.org}

\author{Tianyi Zhang}
\orcid{0000-0002-5468-9347}
\affiliation{%
  \institution{Purdue University}
  \city{West Lafayette}
  \country{USA}
}
\email{tianyi@purdue.edu}

\begin{abstract}
    Large Language Models (LLMs) have recently been widely used for code generation. Due to the complexity and opacity of LLMs, little is known about how these models generate code. We made the first attempt to bridge this knowledge gap by investigating whether LLMs attend to the same parts of a task description as human programmers during code generation. An analysis of six LLMs, including GPT-4, on two popular code generation benchmarks revealed a consistent misalignment between LLMs' and programmers' attention.
We manually analyzed 211 incorrect code snippets and found five attention patterns that can be used to explain many code generation errors. Finally, a user study showed that model attention computed by a perturbation-based method is often favored by human programmers.
Our findings highlight the need for human-aligned LLMs for better interpretability and programmer trust.
\end{abstract}


\maketitle

% --------------------INTRODUCTION----------------------
\section{INTRODUCTION}
\label{sec:intro}
Large Language Models (LLMs) have made significant process on code generation in recent years~\cite{chen2021codex, ahmad2021unified, elnaggar2021codetrans, feng2020codebert, guo2020graphcodebert, jain2021contrastive, fried2022incoder, gpt-j, gpt-neox-library}. 
A recent study~\cite{bubeck2023sparks} shows that GPT-4, the state-of-the-art LLM with 1.7 trillion parameters, can correctly solve 84\%
of the Python programming tasks from the HuamnEval~\cite{chen2021codex} benchmark. 
Despite this great progress, it remains unclear why and how LLMs can generate correct code from natural language descriptions.

Model attention analysis is a common methodology to understand how a model works. It has been widely adopted in computer vision~\cite{huang2021attributes, zhao2020double, fong2018using, jia2018biometric, melicio2018object, nunes2020learning, stocco2022thirdeye, hazard2022importance} to investigate whether a model pays attention to the salient parts of an image when making a decision. In particular, recent studies find that aligning model attention with human attention can effectively enhance model performance~\cite{huang2021attributes, gao2022aligning, bansal2023towards}. 
For instance, Huang et al.~\cite{huang2021attributes} show that the performance of Conv-4-based image classification models can be increased by up to 23\% when they are trained to align with human attention. %(i.e., attending important areas in an image labeled by human annotators). 
Furthermore, previous studies also suggest that users have more confidence and trust in human-aligned models~\cite{boggust2022shared, hendrycks2020aligning, shin2021effects, weitz2019you}.
For instance, Boggust et al.~\cite{boggust2022shared} find that users determine the trustworthiness of a model by checking whether the model makes predictions based on features they consider important.

These findings lead to an important scientific question for LLM-based code generation---{\em do LLMs attend to similar parts of a task description like human programmers in code generation?} \mr{We choose to compare model attention with human attention, since it can help us determine whether LLMs grasp the deep semantics in a task description like humans or whether they just learn superficial patterns from training data, which is known as a common problem in machine learning. Furthermore, by comparing human and model attention patterns, we seek to investigate whether the attention differences can be used to explain some code generation errors and inform new opportunities to improve LLMs for code generation.}
%While recent studies~\cite{paltenghi2021thinking, bansal2023towards, rabin2021understanding} have performed attention analysis of neural models for code summarization, program repair, and method name prediction, they do not look into LLMs or code generation tasks. Furthermore, due to the complexity in the attention mechanism of LLMs (e.g., multiple attention layers and each layer with multiple attention heads), there is still no consensus on how to calculate model attention for LLMs. For instance, some  studies~\cite{wan2022they, liu2023reliability} sum up the attention from all attention heads in all layers and argue that the summation reflects long-distance token relationships, while other work~\cite{zeng2022extensive, bensemann2022eye} extracts attention from the first few layers only and argues that the first few layers represent the syntactic patterns learned by the model.

To bridge the knowledge gap, we made the first effort to unveil the code generation process of LLMs by analyzing \textit{which parts of human language an LLM attends to when generating code}. 
We present a large-scale study that examines the attention alignment between six LLMs and human programmers on two popular code generation benchmarks---
OpenAI's HumanEval benchmark~\cite{chen2021codex} and Google's MBPP benchmark~\cite{austin2021program}. The hypothesis is that LLMs should generate code based on salient words from an NL description similar to human programmers, rather than generating code based on trivial tokens such as prepositions and delimiters. Specifically, we investigated the following research questions in this study:

\begin{itemize}[leftmargin=10mm]
    \setlength\itemsep{0.5mm}
    \item[RQ1] To what extent is model attention aligned with human attention?
    \item[RQ2] Can attention explain errors of code generation models?
    \item[RQ3] What is the impact of different attention calculation methods on attention alignment?
    \item[RQ4] Which attention calculation method is most preferred by programmers?
\end{itemize}

Since none of the existing code generation benchmarks contain programmer attention information,%(i.e., which words or phrases a programmer pays attention to when writing code), 
we created the first programmer attention dataset for the programming tasks from HumanEval and MBPP (1,138 tasks in total). %We chose these two datasets because they are widely used benchmarks and they contain test cases to assess the functional correctness of generated code.
We captured programmers' attention by asking two experienced programmers to manually label words and phrases that they considered essential to solving each programming task. Their labels are validated by a third programmer. % who independently labeled 164 programming tasks and achieved a substantial agreement with the previous two programmers. 
On the other hand, to capture model attention, we implemented and experimented with twelve different attention calculation methods in three categories---{\em self-attention-based}, {\em gradient-based}, and {\em perturbation-based}. To ensure our findings generalize across different LLMs, we analyzed the attention of six LLMs with different sizes, including GPT-4~\cite{chatgpt}, InCoder-1.3B~\cite{fried2022incoder}, CoderGen-2.7B~\cite{nijkamp2022codegen}, PolyCoder-2.7B~\cite{xu2022systematic}, CodeParrot-1.5B~\cite{codeparrot}, and GPT-J-6B~\cite{gpt-j}.

Our study reveals several important insights into the code generation process of LLMs. First, we find a consistent \textit{attention misalignment} in all six LLMs, regardless of the attention calculation methods.
Furthermore, \mr{we performed an in-depth analysis of the attention patterns of 211 incorrect codes generated by the two best models in our study, CodeGen-2.7B, and GPT-4. We found that 27\% of the code generation errors could be explained by five attention patterns.} Finally, perturbation-based methods generated attention scores that are overall more aligned with human attention than other methods. They are also preferred by human programmers, according to a user study of 22 participants. Our findings highlight the need to develop human-aligned LLMs and provide practical guidelines for improving LLM-based code generation and calculating model attention.

In summary, this paper makes the following contributions:
\begin{itemize}
    \item We conducted the first empirical study on the attention alignment of LLMs and human programmers on code generation tasks.
    %\item Our study highlights the importance of human-aligned LLMs and identifies opportunities for improving code generation models through attention analysis.
    \item We conducted a comparative analysis of different attention calculation methods for code generation models through both quantitative experiments and a user study. %Our results indicate that a gradient-based method achieves the best performance in quantitative experiments, while participants in the user study often prefer another perturbation-based method.
    \item \sloppy We made publicly available the first programmer attention dataset of 1,138 Python tasks, which can be used to develop new human-aligned models and evaluate interpretability methods for code generation models. Our code and data have been made available in our GitHub repository~\cite{attention_alignment_empirical_study}.
\end{itemize}
% ------------------------------------------------------

% ---------------------MOTIVATION-----------------------
\section{MOTIVATION AND PRELIMINARIES}
In this section, we first explain the motivation for performing attention analysis for code generation models. Then, we introduce popular benchmarks and metrics for code generation. Finally, we define model attention and describe different kinds of attention calculation methods for LLMs.

\subsection{Motivation}
Model attention analysis is a common task in several domains, such as computer vision~\cite{fong2018using, jia2018biometric, melicio2018object, nunes2020learning}, neural machine translation~\cite{chen2016guided}, and autonomous driving~\cite{stocco2022thirdeye, hazard2022importance, kotseruba2016joint}. Specifically, several studies show that aligning model attention with human attention during the training process can significantly improve the model performance~\cite{huang2021attributes, gao2022aligning, bansal2023towards}. For instance, Huang et al.~\cite{huang2021attributes} show that by aligning the attention of Conv-4-based and ResNet-based models to human attention on images, the performance of these models on image classification can be increased by up to 23\% in the one-shot setting and 10\% in the five-shot setting.
In autonomous driving, many studies have demonstrated that adding attention alignment constraints can help the autonomous driving system to drive safer~\cite{stocco2022thirdeye, hazard2022importance, kotseruba2016joint}. This motivates us to investigate the attention alignment between LLMs and human programmers in the domain of code generation. 

Several recent studies analyzed the attention of neural models for code summarization, program repair, and method name prediction~\cite{paltenghi2021thinking, bansal2023towards, rabin2021understanding}.
Paltenghi et al.~\cite{paltenghi2021thinking} studied the attention alignment between neural model attention and programmer attention on code summarization. %However, their study focuses on CNN and a transformer model trained from scratch instead of LLMs.
Bansal et al.~\cite{bansal2023towards} showed that aligning the attention of neural code summarization models with human attention can effectively improve model performance.
Rabin et al.~\cite{rabin2021understanding} found that pre-trained code models rely heavily on just a few syntactic features in the prompts to perform method name prediction and variable misuse detection.
To the best of our knowledge, none of the existing studies have investigated code generation tasks or LLMs with billions of parameters. Our study fills this gap by analyzing the attention patterns of six LLMs in code generation tasks. 

Finally, another motivation behind this study is the fact that there is still no consensus on how to calculate the attention score of LLM-based code models. This is largely attributed to the complexity of the multi-head, multi-layer attention mechanism adopted by these models. For example, some studies only considered model attention from the first transformer layer and stated that the first layer encodes lexical-level information~\cite{zeng2022extensive, bensemann2022eye}, while other studies summed up the attention from all transformer layers to incorporate long-distance token relationships~\cite{wan2022they, liu2023reliability}. To bridge this gap, we conducted the first comprehensive study on 12 attention calculation methods and systemically evaluated them with quantitative experiments and a user study with 22 participants.

% \mr{In practice, most LLMs today can only correctly solve 30\%-60\% tasks in HumanEval. Several studies~\cite{barke2022grounded, bird2022taking, srikant2021generating} reveal that it takes significant effort to review, debug, and fix incorrect code, which diminishes developer productivity and trust in LLM code generation. So it is important to understand the inner workings of LLMs and why they generate incorrect code.} 

\subsection{Code Generation Benchmarks and Metrics}

\begin{figure}[!t]
    \centering
    \includegraphics[width=0.6\linewidth]{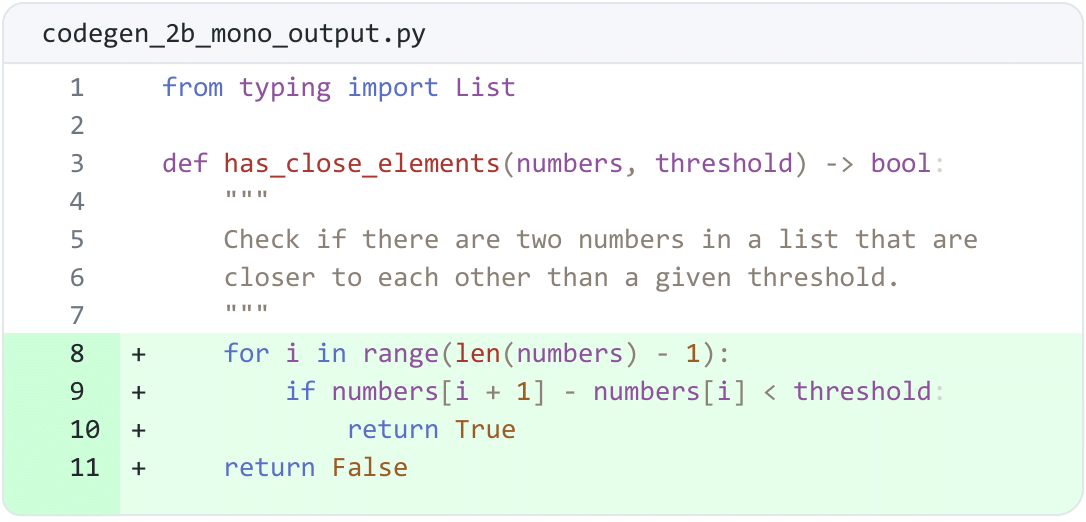}
    \caption{A Python function generated by CodeGen-2.7B~\cite{nijkamp2022codegen}. The generated code is highlighted in \hlgreen{green}.}
    \vspace{-10pt}
    \label{fig:code_generation_example}
\end{figure}

In this work, we focus on code generation tasks that generate a function from a natural language description. Since the breakthrough of OpenAI's Codex model in 2021~\cite{chen2021codex}, this kind of code generation task has become increasingly popular in the research community. 
In this task setting, given a function header and a task description in natural language, an LLM is expected to complete the function according to the task description. 
Figure~\ref{fig:code_generation_example} shows an example.

Code generation models are often evaluated on crowd-sourced programming benchmarks. 
OpenAI developed a programming benchmark called HumanEval and used it to evaluate the original Codex model and its variants~\cite{chen2021codex}.   HumanEval~\cite{chen2021codex} includes 164 Python programming tasks and ground-truth solutions. It is by far the most popular benchmark and has been used to evaluate most LLM-based code generation models. Besides, MBPP is a large benchmark~\cite{austin2021program} with 974 crowd-sourced programming tasks and solutions. Both HumanEval and MBPP include test cases to evaluate the functional correctness of generated code. %CodeXGLUE is another large benchmark that not only covers code generation tasks but also code translation and code summarization tasks~\cite{lu2021codexglue}. However, CodeXGLUE does not include any test cases. 

Two types of metrics are often used to evaluate the performance of LLM-based code generation models. First, if a code generation benchmark includes test cases, one can simply run the test cases to evaluate the correctness of the generated code. A typical metric in this category is \textit{Pass@k}. \textit{Pass@k} was initially introduced by Kulal et al.~\cite{kulal2019spoc}. It measures the percentage of correctly solved programming tasks where $k$ refers to the number of code samples generated by the LLM. If any of the $k$ samples pass all test cases in a task, the task is considered correctly solved. OpenAI then introduces an unbiased version of \textit{Pass@k} to reduce variances, which is widely used to evaluate code generation models these days~\cite{chen2021codex}. %For example, the Codex model achieves 28.8\% in \textit{Pass@1} and 70.2\% in \textit{Pass@100}. 
In practice, many programming tasks do not have existing test cases and some code solutions do not have a well-defined function interface for testing, e.g., a single line of code without clear input and output. Thus, prior work also measures the similarity between generated code and a ground-truth solution as a proxy for correctness. BLEU~\cite{papineni2002bleu} and  CodeBLEU~\cite{ren2020codebleu} are commonly adopted similarity metrics.  BLEU is a metric commonly used to evaluate the quality of machine-generated text. It evaluates the quality of machine-generated text by comparing the presence of n-grams in the generated text to those in reference texts. 
BLEU calculates a score between 0 and 1, with 1 indicating perfect similarity. 
CodeBLEU is designed to adapt BLEU specifically for evaluating generated code. While BLEU primarily considers word-level similarity, CodeBLEU considers the structure and correctness of the generated code, which are crucial aspects in code generation tasks.

\subsection{Model Attention}
In this work, we use {\em model attention} to refer to how important a token in a natural language task description is considered by an LLM during code generation. It implies which parts of the input the model ``attends'' to during code generation. This idea resembles {\em feature importance}~\cite{hooker2018evaluating}, {\em saliency map}~\cite{niebur2007saliency}, and {\em feature attribution}~\cite{molnar2020interpretable} in the XAI literature. We describe three kinds of model attention calculation methods as follows.

\subsubsection{Self-attention-based Methods.} 
\label{sec:self_attention_based_methods}
The self-attention mechanism allows transformers to weigh the importance of different parts of the input when making predictions. 
By focusing on the most relevant tokens with the highest self-attention scores, the transformers can make better predictions by capturing relationships and dependencies in the input.

\begin{figure}[!t]
    \centering
\includegraphics[width=.6\linewidth]{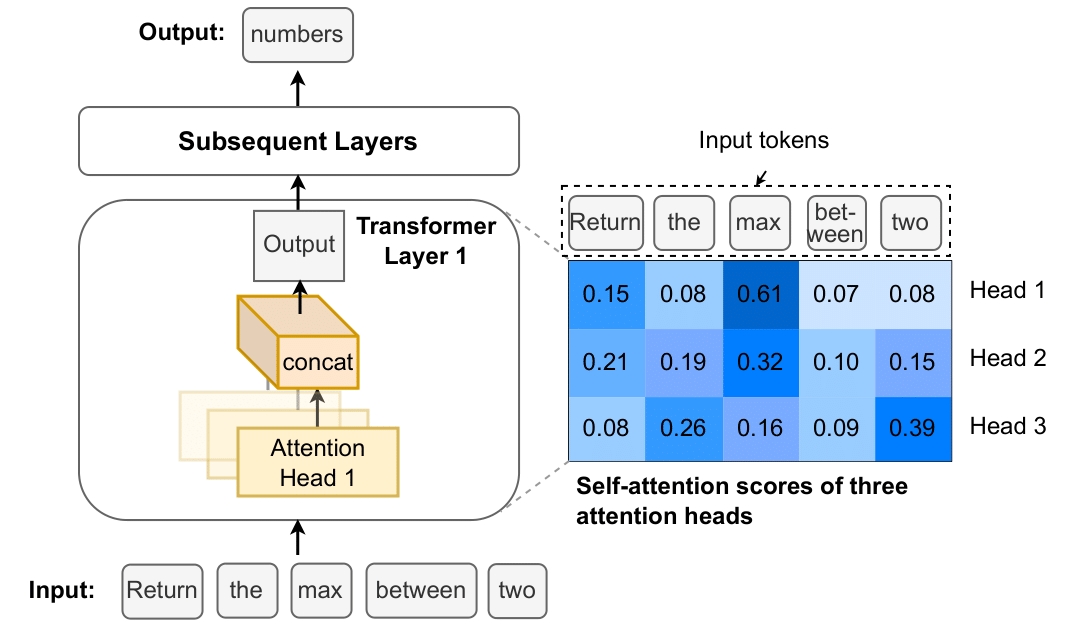}
    \caption{Attention matrix of the first attention head of the transformer layer in CodeGen-2.7B}
    \vspace{-10pt}
    \label{fig:attention_matrix}
\end{figure}

An LLM includes multiple \textit{transformer layers}, each of which includes multiple \textit{attention heads}. These attention heads independently calculate self-attention scores between different input tokens. 
Therefore, a transformer model could have multiple sources of model attention from different transformer layers and different attention heads in each layer.
For example, Figure~\ref{fig:attention_matrix} shows the self-attention scores of the first three attention heads of the first transformer layer in CodeGen-2.7B when generating token \textit{``numbers''} from input sequence \textit{``Return the max between two''}. The self-attention scores calculated by different attention heads differ for the same token. Different heads represent different kinds of ``focus'' from the model.

To calculate model attention on the input sequence, vectors of self-attention scores from different attention layers and different attention heads in each layer need to be aggregated into a single vector to represent the overall importance of each input token to the model prediction. 
However, although self-attention scores have been generally used as model attention~\cite{clark2019does, galassi2020attention, li2016understanding, vashishth2019attention}, there is still no consensus on how to aggregate self-attention scores from different layers and attention heads.
For example, some studies sum up the attention scores attention from all transformer layers and all attention heads in each layer~\cite{zhang2022does} as the final attention scores.  These studies argue that this summation strategy can capture long-distance token relationships. Some other studies only use the attention scores from the first transformer layer and argue that the first layer captures lexical-level dependencies~\cite{zeng2022extensive, bensemann2022eye}. 

To reveal how different ways to aggregate self-attention affect the alignment between model and human attention, we experimented with six self-attention-based methods in this study (detailed in Section 4.2.1).

\subsubsection{Gradient-based Methods.}
Gradient-based methods leverage the gradients of the model's predictions concerning the input features to calculate the model's attention. The calculation of gradient-based methods includes two different steps: (1) perform a forward pass of input of interest, and (2) calculate gradients using backpropagation through the neural network's layers. By analyzing the magnitudes of these gradients, these methods can identify which input tokens are most influential in determining the output.
For example, {\em Integrated Gradients} is a gradient-based method that computes the integral of the gradients of the model's output concerning each input feature~\cite{sundararajan2017axiomatic, denil2014extraction, shrikumar2017learning}.
In this study, we experimented with two gradient-based methods used in previous work~\cite{simonyan2013deep, shrikumar2017learning} with details in Section 4.2.2.

\subsubsection{Perturbation-based Methods.}
\sloppy Different from the previous two categories of methods, perturbation-based methods~\cite{wu2020perturbed, vashishth2019attention} are model-agnostic. In other words, they do not require access to the internal information of a model. Perturbation-based methods are particularly useful for calculating the attention of commercial models such as GPT-4, since these models do not reveal their self-attention layers or gradients to users. 

Perturbation-based methods first mutate the input and then calculate the model's attention based on the output differences.
LIME~\cite{ribeiro2016should} and SHAP~\cite{lundberg2017unified} are two popular perturbation-based methods. LIME~\cite{ribeiro2016should} generates a local explanation by approximating the specific model predictions with a simpler model (e.g., a linear classifier). SHAP~\cite{lundberg2017unified} enhances LIME by perturbing the input based on game theory and uses Shapely value to estimate different tokens' importance. 
 
A limitation of these two methods is that they often require a large number of perturbed samples to ensure estimation accuracy. This is costly for calculating the attention for GPT-4, since we need to query GPT-4 many times. Furthermore, LIME and SHAP only mutate an input by deleting tokens, which may significantly change the meaning or the structure of an input. To address this limitation, more recent perturbation-based methods choose to substitute tokens with similar or semantically related tokens in the context~\cite{wu2020perturbed, liu2018nlize}. They often use a masked language model such as BERT~\cite{devlin2018bert} to predict similar or semantically related tokens to substitute existing tokens in an input. Then, they measure the influence of these substitutions on the output. 
In this study, we experimented with SHAP~\cite{lundberg2017unified} and the BERT masking-based method~\cite{wu2020perturbed} (detailed in Section 4.2.3).

% ------------------------------------------------------

% ----------------------DATASET-------------------------
\section{THE CONSTRUCTION OF THE PROGRAMMER ATTENTION DATASET}
\label{sec:dataset}
Since none of the existing code generation benchmarks contain programmer attention information (i.e., which words or phrases a programmer considers important when writing code), we created the first programmer attention dataset based on the 1,138 programming tasks, including all 164 prompts from HumanEval~\cite{chen2021codex} and the 974 prompts from MBPP~\cite{austin2021program}. We selected these two datasets, since they are widely used to evaluate code generation models and they also provide test cases for each programming task, which is important for calculating the correctness of model-generated code.

The first two authors, who have more than five years of programming experience in Python, manually labeled the words and phrases they considered important to solve the programming task in each task description. Before the labeling process, the two labelers went through programming tasks in HumanEval to familiarize themselves with the programming tasks and the code solutions. Then, they first independently labeled the first 20 task descriptions in HumanEval. This first round of labeling had a Cohen's Kappa score of 0.68. The two labelers discussed the disagreements and summarized four kinds of keywords that both of them considered important. The four keyword types are summarized below:

\begin{itemize}
    \item \textbf{Data types}: Words or phrases that describe the types of data that the code should input or output, such as \textit{``string''}, \textit{``number''}, or \textit{``list''}.
    \item \textbf{Operators}: Words or phrases that describe the operations that the code should perform on the data, such as \textit{``compare''}, \textit{``sort''}, \textit{``filter''}, or \textit{``search''}.
    \item \textbf{Conditionals}: Words or phrases that specify the conditions under which the code should execute, such as phrases after \textit{``if''} and \textit{``when''} in a task description. 
    \item \textbf{Properties}: Important properties of the manipulated data and operations, such as quantifiers (e.g., \textit{``all''}, \textit{``one''}), adjectives (e.g., \textit{``first''}, \textit{``closer''}), and adverbs (e.g., \textit{``every''}, \textit{``none''}).
\end{itemize}

\mr{Although there are only four types of keywords, each type is designed to be high-level and inclusive. For instance, the ``operator'' type refers to any kind of operation on the data, such as \textit{``sort a list''}, \textit{``connect a database''}, and \textit{``plot a graph''}.} With this labeling standard, the two labelers proceed to label the remaining 144 task descriptions in the HumanEval dataset. The Cohen's Kappa score of this round of labeling increased to 0.72, indicating a substantial agreement~\cite{mchugh2012interrater}. Then, they discussed and resolved all disagreements. 

To verify these labels, the third author, who was not involved in the previous labeling process, independently labeled the 164 task descriptions from the HumanEval dataset. Since Cohen's Kappa can only calculate the agreement level between two labelers, we used Fleiss' Kappa to measure the agreement between the third labeler and the initial labels from the first two labelers. The Fleiss' Kappa score is 0.64, indicating a substantial agreement~\cite{fleiss1971measuring}. This result shows labels made by the first two labelers are reasonable and can be accepted by other programmers. Then, the first two labelers continued to label 974 programming tasks in the MBPP dataset independently, which resulted in a Cohen's Kappa score of 0.73.  Finally, they resolved all disagreements and used the final set of labels as the programmer's attention dataset. The entire labeling process takes 192 person-hours.

\begin{figure}[!t]
    \centering
    \includegraphics[width=.6\linewidth]{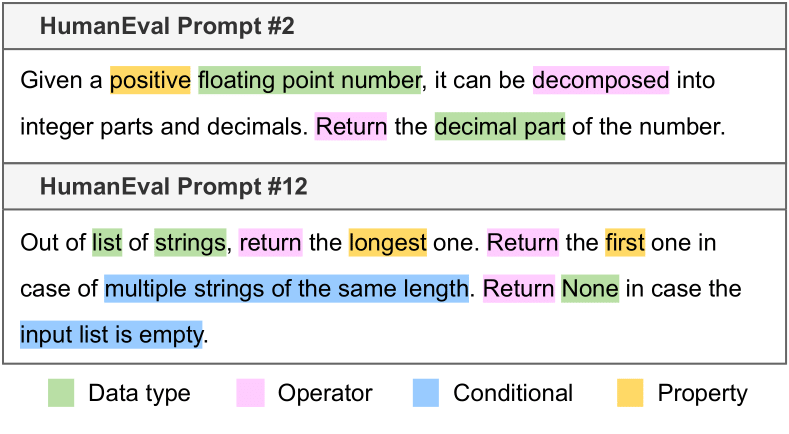}
    \vspace{-10pt}
    \caption{Two examples of labeled prompts from our dataset.}
      \label{fig:description_label}
      \vspace{-10pt}
\end{figure}

On average, each task description has 29.6 words, among which 7 words are considered important by both labelers.  Among all four types of keywords, \textit{property} keywords (45.2\%) are labeled the most frequently by the two labelers, followed by {\em operators} (27.2\%), {\em conditional} (25\%), and {\em data types} (2.6\%).  Figure~\ref{fig:description_label} shows two examples of labeled task descriptions. The four types of keywords are labeled in different colors.

% ------------------------------------------------------

% --------------------METHODOLOGY-----------------------
\section{METHODOLOGY}
This section describes the study design to answer the research questions listed in Section~\ref{sec:intro}.

\subsection{Code Generation Models}
In this study, we select six LLMs with different sizes and different model performances on code generation tasks. Table~\ref{table:model_table} shows the size, the number of self-attention layers, the number of attention heads in each layer, and the model performance on the combined dataset of HumanEval and MBPP in terms of Pass@1.
We describe each model below.

\begin{itemize}
    \item \textbf{InCoder-1.3B~\cite{fried2022incoder}} is an open-source code language model from Meta AI. It is trained on 159 GB of permissively licensed code from GitHub, GitLab, and Stack Overflow. Compared with other LLMs, it adopts a new causal masking objective, which allows it to infill blocks of code conditioned on the arbitrary left and right contexts. We used the largest pre-trained model released by Meta, including 1.3B parameters and 24 transformer layers. 
    \item \textbf{PolyCoder-2.7B~\cite{xu2022systematic}} is an open-source model from CMU. It is based on the GPT-2 architecture and is designed to be the open-source counterpart of OpenAI Codex~\cite{chen2021codex}, since Codex is not open-sourced. It is trained on 249GB of code and has 2.7B parameters.  
    \item \textbf{CodeGen-Mono-2.7B~\cite{nijkamp2022codegen}} is an open-source model from Salesforce Research. It follows a standard transformer autoregressive model architecture with rotary position embedding. 
    \item \textbf{CodeParrot-1.5B~\cite{codeparrot}} is another open-source effort of training a GPT-2 model for code generation. It is trained on 180GB Python Code and has 1.5B parameters. 
    \item \textbf{GPT-J-6B~\cite{gpt-j}} is an open-source model from EleutherAI. It adopts a transformer architecture similar to GPT-3. It is trained on 825 GB text data, which includes 95GB code from GitHub.
        \item \textbf{GPT-4~\cite{chatgpt}} is the state-of-the-art language model developed by OpenAI. Since the internal structure of GPT-4 is not publicly disclosed, we do not include the number of layers and heads of GPT-4 in Table~\ref{table:model_table}. It is reported that GPT-4 has about 1.76 trillion parameters~\cite{gpt4params}. \original{We did not calculate \textit{Pass@10} and \textit{Pass@100} for GPT-4 due to limited budget.} We used the API (\texttt{gpt-4}) provided by OpenAI to query GPT-4.
\end{itemize}

\begin{table}
\caption{Code generation models included in this study.}
\label{table:model_table}
\small        
\resizebox{0.4\linewidth}{!}{
\begin{tabular}{|l|c|c|r|}
\hline
\multicolumn{1}{|l|}{\textbf{Model}} & \multicolumn{1}{c|}{\textbf{Layer}} & \multicolumn{1}{c|}{\textbf{Head}} & \multicolumn{1}{c|}{\textbf{Pass@1}} \\ \hline\hline
\textbf{InCoder-1.3B}               & 24                                  & 32                                 & 15.20\%                              \\
\textbf{PolyCoder-2.7B}             & 32                                  & 32                                 & 5.59\%                               \\
\textbf{CodeGen-2.7B}               & 32                                  & 32                                 & 23.70\%                              \\
\textbf{CodeParrot-1.5B}            & 48                                  & 25                                 & 3.58\%                               \\
\textbf{GPT-J-6B}                   & 28                                  & 16                                 & 11.62\%                              \\
\textbf{GPT-4}                      & -                                   & -                                  & 67\%                                 \\
\hline
\end{tabular}
}
\end{table}

\subsection{Model Attention Calculation}
\label{sec:attention_calculation}
We experimented with twelve attention calculation methods from three different categories: six {\em self-attention-based} methods, four {\em gradient-based} methods, and two  {\em perturbation-based} methods.

\subsubsection{Self-attention-based Methods.} Given that LLMs have multiple attention layers, there is currently no consensus on what is the right way to aggregate those self-attentions to explain LLMs. Zeng et al.~\cite{zeng2022extensive} show that the first attention layer is indicative of which tokens the model attends to, while Wan et al.~\cite{wan2022they} show that deeper attention layers are better at capturing long-distance dependencies and program structure. To perform a comprehensive analysis, we decide to experiment with three settings: (1) only using the first attention layer (denoted as {\em first}), (2) only using the last attention layer (denoted as {\em last}), and (3) using all attention layers (denoted as {\em all}). 

To aggregate self-attentions across different attention heads in a layer, we follow the previous work~\cite{zhang2022does} by summing the attention values from different heads. Finally, since LLMs generate code in an autoregressive manner, their attention changes in each step as they read more tokens from the input and as they generate more code. We are curious about which input tokens the model highly attends to as they read the input and which input tokens the model highly attends to as they generate code. So we consider two experiment settings: (1) summing up the attention scores assigned to each token during the input reading process (denoted as {\em READING}), and (2) summing up the attention scores assigned to each token during the code generation process (denoted as {\em CODING}).
Given the three settings in layer-wise attention aggregation and the two settings in step-wise attention aggregation, we have a total of six experiment settings: \textit{READING\_first}, \textit{CODING\_first}, \textit{READING\_last}, \textit{CODING\_last}, \textit{READING\_all}, and \textit{CODING\_all}.

\subsubsection{Gradient-based Methods.} 
\sloppy We consider two different methods to calculate gradient-based model attention: (1) \textit{Saliency}~\cite{simonyan2013deep}, and (2) \textit{Input$\times$Gradient}~\cite{shrikumar2017learning}. The saliency method calculates the model's attention by computing model gradients with respect to the input. Given a LLM $\mathcal{F}$, suppose an input is $X=[x_1, x_2, \dots, x_n]$, where $n$ is the length of the input. \mr{The attention $s_i$ on $x_i$ is calculated as $s_i = \frac{\partial \mathcal{F}(X)}{\partial x_i}$.}
Different from the saliency method, \textit{Input$\times$Gradient} further multiplies gradients with the input's embedding values. \mr{The attention $s_i$ on $x_i$ is calculated as $s_i = x_i\cdot{\frac{\partial \mathcal{F}(X)}{\partial x_i}}$.}

Similar to self-attentions, gradients also change constantly at each generative step. Thus, we also experimented with the two step-wise attention aggregation settings as in the self-attention methods. The combination of the two gradient calculation methods with the two step-wise aggregation settings results in four gradient-based methods---\textit{Input$\times$Gradient\_reading}, \textit{Input$\times$Gradient\_coding}, \textit{Saliency\_reading}, and \textit{Saliency\_coding}.

\subsubsection{Perturbation-based Methods.} 
\sloppy We consider two different perturbation-based methods: (1) \textit{SHAP}~\cite{lundberg2017unified}, which masks the input through deleting some tokens, and (2) \textit{BERT Masking}~\cite{wu2020perturbed}, which masks the input through substituting a token with its masked language modeling prediction result from BERT.

\begin{itemize}
    \item \textit{SHAP}. We use the official SHAP library with a perturbation count equal to 50 to calculate the model's attention (SHAP scores) on different tokens.\footnote{https://shap.readthedocs.io/en/latest/} We aggregate SHAP scores on predicting different tokens by summing them up. 
    \item \textit{BERT Masking}. Since the code from the original BERT Masking paper~\cite{wu2020perturbed} is not publicly available, we re-implemented this method based on the description in the paper. Specifically, given each token in an input prompt, this method masks it and uses a pre-trained BERT model from HuggingFace to predict the most likely token in the masked position. Then, it substitutes the original token with the predicted token and prompts the LLM to regenerate the code solution. This method then calculates the model's attention on this token by calculating the BLEU score~\cite{papineni2002bleu} between the original code solution and the new solution. We iterate through all tokens in the input prompts to obtain model's attention on different tokens.
\end{itemize}

\subsection{Attention Alignment Measurement}
\label{sec:attention_alignment}
\mr{We measured the attention alignment between models and human programmers using two robust metrics---San Martino's token overlapping metrics~\cite{da2019fine} and Krippendorff’s alpha~\cite{krippendorff2018content}. We also experimented with two simple metrics, Cohen's kappa and keyword coverage rate, and obtained similar results. Due to the page limit, we only reported the results of the first two metrics in this paper. The results of the other two metrics are included in the GitHub repository~\cite{attention_alignment_empirical_study}.}

\mr{\subsubsection{San Martino’s Token Overlapping Metrics~\cite{da2019fine}} calculate the overlap between two sets of tokens in terms of precision, recall, and F-1 score. It was initially designed for sequence labeling tasks in NLP~\cite{da2019fine} and has been recently adopted in SE studies as a robust metric for toxicity detection in code review comments~\cite{sarker2023toxispanse}. Specifically, it assigns partial credit for partial overlaps between two sets of tokens, which makes it a suitable choice for our task. Thus, we follow \cite{da2019fine} to compare salient words selected by humans and models. For human attention, a token is considered {\em highly attended} if it is labeled as an important word by human programmers, as described in Section~\ref{sec:dataset}. For model attention, since the attention calculation methods in Section~\ref{sec:attention_calculation} compute a continuous score for each token, it is hard to determine a universal threshold to decide which token is highly attended by a model. Thus, we rank the tokens in a programming task description based on their attention scores and select the top $K$ tokens as the {\em highly attended} tokens by a model. Since the average of human-labeled important words is 7 per task description, we experiment with $K=5, 10, 20$ in our study. Given a set of highly attended tokens by human programmers and a set of highly attended tokens by a model, we follow the equations in~\cite{sarker2023toxispanse} to compute precision, recall, and F-1 score.}
We report all three scores in Table~\ref{table:overall_result}.

\subsubsection{Krippendorff’s Alpha~\cite{krippendorff2004reliability}}
\begin{figure}[t]
    \centering
    \includegraphics[width=0.45\linewidth]{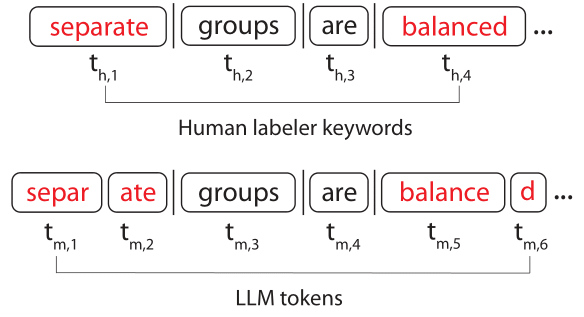}
    \caption{Mapping NL words to LLM sub-tokens}
    \vspace{-10pt}
    \label{fig:tokens}
\end{figure}

\mr{is a robust statistical measure for inter-labeler agreement, where $\alpha=1$ indicates perfect agreement, $\alpha=0$ indicates no agreement, and $\alpha=-1$ indicates complete disagreement.
It can be used to measure the agreement level between human programmers and LLMs on important words in a programming task description. 
Compared to other inter-rater agreement metrics such as Cohen's kappa~\cite{cohen1960coefficient}, \ka{} is more robust to the number of coders, missing data, and sample size. To calculate \ka{}, human and model labels must be stored in vectors of the same length. This is hard because LLMs perform subword tokenization to handle out-of-vocabulary words with a manageable vocabulary size. For example, in Figure~\ref{fig:tokens}, the word ``separate'' is tokenized into two tokens: ``separ'' and ``ate'' through byte pair encoding (BPE). During inference, an LLM will compute attention scores separately for these two subtokens, though they are from the same English word.} 

To address this challenge, we map NL words back to LLM tokens. If the model tokenizes a natural language word into multiple tokens, all sub-tokens will be considered selected by human labelers. Using the same example from Figure~\ref{fig:tokens}, if the natural language word ``separate'' is selected by the human labelers, both sub-tokens ``separ'' and ``ate'' will be considered selected and represented by 1 in the vectors used to calculate \ka{}.

%\mr{Both \ka{} and \sm{} do not consider the ordering of tokens. This is because model attention analysis is typically formulated as whether a token is attended or to what extent a token is attended. The attention mechanism used in LLMs treats each token independently without considering their ordering. Instead, LLMs use positional encoding to track the ordering of tokens, which is out of scope for attention analysis.}

\subsection{User Study Design}
\label{sec:user-study-design}
To answer RQ4, we conducted a user study to evaluate the different attention calculation methods.\footnote{Our user study questionnaire and participants' responses are available in our GitHub repository: \href{\website}{\website}.} We recruited 22 students (18 males and 4 females) through the department mailing lists in a CS department. According to our user study, all participants have an average of 5.62 years of programming experience. All participants have some basic understanding of model attention in machine learning.

We randomly selected 8 task descriptions from our dataset. For each task, we leveraged CodeGen-2.7B, the best-performing open-source model on HumanEval in our experiment, to calculate its model attention. We did not consider GPT-4 in this user study, since it is close-sourced and we cannot compute its attention using self-attention-based methods and gradient-based methods. We selected one attention calculation method from each category (\textit{self-attention-based}, \textit{perturbation-based}, and \textit{gradient-based}): $CODING\_{last}$, ${Input\times Gradient\_{coding}}$, and \textit{SHAP}.

In each user study, participants first read the task description to understand the programming task and then read the code generated by CodeGen. For each attention method, we render a highlighted version of the task description similar to  Figure~\ref{fig:description_label}, where the top 10 attended tokens are highlighted based on the attention scores computed by this method. We chose to render the top 10 important keywords since it is close to the average number of important words (7) labeled in our dataset. Participants were then asked to rate each attention method by indicating their agreement with the following three statements on a 7-point Likert scale (1---completely disagree, 7---completely agree).

\begin{enumerate}[label={(\arabic*)}]
    \item[Q1] \textit{The model-attended keywords align with my attention when reading the natural language task description.}
    \item[Q2] \textit{This attention explains why the model succeeded in or failed at generating the correct code.}
    \item[Q3] \textit{I want to see this attention when working with code generation models in real life.}
\end{enumerate}

\begin{table*}[t]
    \caption{Human-model attention alignment calculated by different methods in terms of San Martino's Precision (P), Recall (R), F1 score (F1), and Krippendorff's alpha score (KA). Among the 12 different attention calculation methods, the one that produces the most aligned result for each model under each $K$ setting in each metric is highlighted in \yellow{yellow}. Note that we only experimented with \mr{perturbation-based methods} on GPT-4, since GPT-4 does not provide access to its self-attention layers and gradients.}
    \label{table:overall_result}
\begin{minipage}{\linewidth}
\begin{subtable}[h]{\textwidth}
    \setlength{\tabcolsep}{1.5pt}

    \caption{Perturbation-based methods.}
    \label{table:pertubation_based}
    \scriptsize
    \resizebox{\linewidth}{!}{

\begin{tabular}{|l|rrrrrrrrrrrr|rrrrrrrrrrrr|}
\hline
                                  & \multicolumn{12}{c|}{\textbf{BERT\_masking}}                                                                                                                                                                                                                                                                                                                                                                                        & \multicolumn{12}{c|}{\textbf{SHAP}}                                                                                                                                                                                                                                                                                                                                                                                                                  \\ \cline{2-25} 
                                  & \multicolumn{4}{c|}{\textbf{Top 5}}                                                                                                                   & \multicolumn{4}{c|}{\textbf{Top 10}}                                                                                                 & \multicolumn{4}{c|}{\textbf{Top 20}}                                                                                                 & \multicolumn{4}{c|}{\textbf{Top 5}}                                                                                                                   & \multicolumn{4}{c|}{\textbf{Top 10}}                                                                                                                  & \multicolumn{4}{c|}{\textbf{Top 20}}                                                                                                 \\ \cline{2-25} 
\multirow{-3}{*}{\textbf{Method}} & \multicolumn{1}{c}{\textbf{P}} & \multicolumn{1}{c}{\textbf{R}} & \multicolumn{1}{c}{\textbf{F1}} & \multicolumn{1}{c|}{\textbf{KA}}                  & \multicolumn{1}{c}{\textbf{P}} & \multicolumn{1}{c}{\textbf{R}} & \multicolumn{1}{c}{\textbf{F1}} & \multicolumn{1}{c|}{\textbf{KA}} & \multicolumn{1}{c}{\textbf{P}} & \multicolumn{1}{c}{\textbf{R}} & \multicolumn{1}{c}{\textbf{F1}} & \multicolumn{1}{c|}{\textbf{KA}} & \multicolumn{1}{c}{\textbf{P}} & \multicolumn{1}{c}{\textbf{R}} & \multicolumn{1}{c}{\textbf{F1}} & \multicolumn{1}{c|}{\textbf{KA}}                  & \multicolumn{1}{c}{\textbf{P}} & \multicolumn{1}{c}{\textbf{R}} & \multicolumn{1}{c}{\textbf{F1}} & \multicolumn{1}{c|}{\textbf{KA}}                  & \multicolumn{1}{c}{\textbf{P}} & \multicolumn{1}{c}{\textbf{R}} & \multicolumn{1}{c}{\textbf{F1}} & \multicolumn{1}{c|}{\textbf{KA}} \\ \hline
\textbf{Incoder}                  & \cellcolor[HTML]{FCCD7D}48.8\% & \cellcolor[HTML]{FCCD7D}40.6\% & \cellcolor[HTML]{FCCD7D}44.3\%  & \multicolumn{1}{r|}{\cellcolor[HTML]{FCCD7D}0.25} & 40.8\%                         & 67.4\%                         & 50.8\%                          & \multicolumn{1}{r|}{0.25}        & 36.6\%                         & \cellcolor[HTML]{FCCD7D}90.3\% & 52.1\%                          & 0.12                             & 33.9\%                         & 28.4\%                         & 30.9\%                          & \multicolumn{1}{r|}{0.11}                         & 33.4\%                         & 55.9\%                         & 41.8\%                          & \multicolumn{1}{r|}{0.18}                         & 32.8\%                         & 86.7\%                         & 47.6\%                          & 0.15                             \\ \cline{1-1}
\textbf{CodeGen}                  & \cellcolor[HTML]{FCCD7D}51.8\% & \cellcolor[HTML]{FCCD7D}43.2\% & \cellcolor[HTML]{FCCD7D}47.1\%  & \multicolumn{1}{r|}{\cellcolor[HTML]{FCCD7D}0.29} & \cellcolor[HTML]{FCCD7D}41.8\% & \cellcolor[HTML]{FCCD7D}68.9\% & \cellcolor[HTML]{FCCD7D}52.0\%  & \multicolumn{1}{r|}{0.27}        & \cellcolor[HTML]{FCCD7D}36.8\% & \cellcolor[HTML]{FCCD7D}90.7\% & \cellcolor[HTML]{FCCD7D}52.4\%  & 0.13                             & 33.0\%                         & 27.5\%                         & 30.0\%                          & \multicolumn{1}{r|}{0.10}                         & 32.9\%                         & 54.9\%                         & 41.1\%                          & \multicolumn{1}{r|}{0.17}                         & 33.0\%                         & 87.1\%                         & 47.9\%                          & 0.16                             \\ \cline{1-1}
\textbf{CodeParrot}               & 51.5\%                         & 43.0\%                         & 46.9\%                          & \multicolumn{1}{r|}{0.29}                         & 42.1\%                         & 69.2\%                         & 52.3\%                          & \multicolumn{1}{r|}{0.28}        & 36.6\%                         & \cellcolor[HTML]{FCCD7D}90.2\% & 52.0\%                          & 0.12                             & 33.2\%                         & 27.7\%                         & 30.2\%                          & \multicolumn{1}{r|}{0.10}                         & 32.7\%                         & 54.7\%                         & 40.9\%                          & \multicolumn{1}{r|}{0.17}                         & 33.3\%                         & 87.6\%                         & 48.3\%                          & 0.17                             \\ \cline{1-1}
\textbf{GPT-J-6B}                 & \cellcolor[HTML]{FCCD7D}49.9\% & \cellcolor[HTML]{FCCD7D}41.3\% & \cellcolor[HTML]{FCCD7D}45.2\%  & \multicolumn{1}{r|}{\cellcolor[HTML]{FCCD7D}0.26} & \cellcolor[HTML]{FCCD7D}41.0\% & \cellcolor[HTML]{FCCD7D}67.4\% & \cellcolor[HTML]{FCCD7D}51.0\%  & \multicolumn{1}{r|}{0.26}        & \cellcolor[HTML]{FCCD7D}36.6\% & \cellcolor[HTML]{FCCD7D}90.3\% & \cellcolor[HTML]{FCCD7D}52.1\%  & 0.12                             & 33.3\%                         & 27.8\%                         & 30.3\%                          & \multicolumn{1}{r|}{0.10}                         & 33.3\%                         & 55.7\%                         & 41.7\%                          & \multicolumn{1}{r|}{0.18}                         & 33.1\%                         & 87.1\%                         & 47.9\%                          & 0.16                             \\ \cline{1-1}
\textbf{PolyCoder}                & \cellcolor[HTML]{FCCD7D}49.3\% & \cellcolor[HTML]{FCCD7D}41.3\% & \cellcolor[HTML]{FCCD7D}45.0\%  & \multicolumn{1}{r|}{\cellcolor[HTML]{FCCD7D}0.26} & 40.9\%                         & \cellcolor[HTML]{FCCD7D}67.7\% & 51.0\%                          & \multicolumn{1}{r|}{0.26}        & \cellcolor[HTML]{FCCD7D}36.6\% & \cellcolor[HTML]{FCCD7D}90.4\% & \cellcolor[HTML]{FCCD7D}52.1\%  & 0.12                             & 34.4\%                         & 28.9\%                         & 31.4\%                          & \multicolumn{1}{r|}{0.12}                         & 33.5\%                         & 56.2\%                         & 42.0\%                          & \multicolumn{1}{r|}{0.19}                         & 33.2\%                         & 87.2\%                         & 48.1\%                          & 0.16                             \\ \cline{1-1}
\textbf{GPT-4}                    & 32.7\%                         & 27.6\%                         & 30.0\%                          & \multicolumn{1}{r|}{0.04}                         & \cellcolor[HTML]{FCCD7D}36.7\% & \cellcolor[HTML]{FCCD7D}61.7\% & \cellcolor[HTML]{FCCD7D}46.0\%  & \multicolumn{1}{r|}{0.17}        & \cellcolor[HTML]{FCCD7D}36.3\% & \cellcolor[HTML]{FCCD7D}89.4\% & \cellcolor[HTML]{FCCD7D}51.7\%  & 0.11                             & \cellcolor[HTML]{FCCD7D}34.7\% & \cellcolor[HTML]{FCCD7D}29.2\% & \cellcolor[HTML]{FCCD7D}31.7\%  & \multicolumn{1}{r|}{\cellcolor[HTML]{FCCD7D}0.12} & 34.2\%                         & 57.4\%                         & 42.9\%                          & \multicolumn{1}{r|}{\cellcolor[HTML]{FCCD7D}0.20} & 33.1\%                         & 87.0\%                         & 47.9\%                          & \cellcolor[HTML]{FCCD7D}0.16     \\ \hline
\end{tabular}

}
\end{subtable}
\vspace{1mm}
\vfill
\begin{subtable}[h]{\textwidth}
\setlength{\tabcolsep}{1.5pt}

    \caption{Gradient-based methods}
    \label{table:gradient_based}
    \scriptsize
    \resizebox{\linewidth}{!}{

\begin{tabular}{|l|rrrrrrrrrrrr|rrrrrrrrrrrr|}
\hline
                                  & \multicolumn{12}{c|}{\textbf{Input$\times$Gradient\_reading}}                                                                                                                                                                                                                                                                                                                                                      & \multicolumn{12}{c|}{\textbf{Input$\times$Gradient\_coding}}                                                                                                                                                                                                                                                                                                                                                                        \\ \cline{2-25} 
                                  & \multicolumn{4}{c|}{\textbf{Top 5}}                                                                                                  & \multicolumn{4}{c|}{\textbf{Top 10}}                                                                                                 & \multicolumn{4}{c|}{\textbf{Top 20}}                                                                                                 & \multicolumn{4}{c|}{\textbf{Top 5}}                                                                                                  & \multicolumn{4}{c|}{\textbf{Top 10}}                                                                                                                  & \multicolumn{4}{c|}{\textbf{Top 20}}                                                                                                 \\ \cline{2-25} 
\multirow{-3}{*}{\textbf{Method}} & \multicolumn{1}{c}{\textbf{P}} & \multicolumn{1}{c}{\textbf{R}} & \multicolumn{1}{c}{\textbf{F1}} & \multicolumn{1}{c|}{\textbf{KA}} & \multicolumn{1}{c}{\textbf{P}} & \multicolumn{1}{c}{\textbf{R}} & \multicolumn{1}{c}{\textbf{F1}} & \multicolumn{1}{c|}{\textbf{KA}} & \multicolumn{1}{c}{\textbf{P}} & \multicolumn{1}{c}{\textbf{R}} & \multicolumn{1}{c}{\textbf{F1}} & \multicolumn{1}{c|}{\textbf{KA}} & \multicolumn{1}{c}{\textbf{P}} & \multicolumn{1}{c}{\textbf{R}} & \multicolumn{1}{c}{\textbf{F1}} & \multicolumn{1}{c|}{\textbf{KA}} & \multicolumn{1}{c}{\textbf{P}} & \multicolumn{1}{c}{\textbf{R}} & \multicolumn{1}{c}{\textbf{F1}} & \multicolumn{1}{c|}{\textbf{KA}}                  & \multicolumn{1}{c}{\textbf{P}} & \multicolumn{1}{c}{\textbf{R}} & \multicolumn{1}{c}{\textbf{F1}} & \multicolumn{1}{c|}{\textbf{KA}} \\ \hline
\textbf{Incoder}                  & 33.7\%                         & 25.9\%                         & 29.3\%                          & \multicolumn{1}{r|}{0.13}        & 36.9\%                         & 46.8\%                         & 41.2\%                          & \multicolumn{1}{r|}{0.32}        & 35.6\%                         & 52.7\%                         & 42.5\%                          & 0.28                             & 34.6\%                         & 27.5\%                         & 30.7\%                          & \multicolumn{1}{r|}{0.14}        & 37.2\%                         & 47.1\%                         & 41.6\%                          & \multicolumn{1}{r|}{0.33}                         & 35.6\%                         & 52.8\%                         & 42.5\%                          & 0.28                             \\ \cline{1-1}
\textbf{CodeGen}                  & 44.8\%                         & 34.2\%                         & 38.8\%                          & \multicolumn{1}{r|}{0.22}        & 39.4\%                         & 59.8\%                         & 47.5\%                          & \multicolumn{1}{r|}{0.28}        & 35.2\%                         & 87.3\%                         & 50.1\%                          & 0.22                             & 46.0\%                         & 34.7\%                         & 39.5\%                          & \multicolumn{1}{r|}{0.23}        & 41.0\%                         & 61.5\%                         & 49.2\%                          & \multicolumn{1}{r|}{\cellcolor[HTML]{FCCD7D}0.31} & 36.2\%                         & 88.5\%                         & 51.4\%                          & \cellcolor[HTML]{FCCD7D}0.25     \\ \cline{1-1}
\textbf{CodeParrot}               & 55.0\%                         & 43.2\%                         & 48.4\%                          & \multicolumn{1}{r|}{0.33}        & 45.8\%                         & 71.2\%                         & 55.8\%                          & \multicolumn{1}{r|}{0.40}        & 35.1\%                         & 87.5\%                         & 50.1\%                          & 0.22                             & 57.9\%                         & 44.9\%                         & 50.6\%                          & \multicolumn{1}{r|}{0.37}        & 46.8\%                         & 71.8\%                         & 56.7\%                          & \multicolumn{1}{r|}{0.42}                         & 36.0\%                         & 88.7\%                         & 51.2\%                          & 0.24                             \\ \cline{1-1}
\textbf{GPT-J-6B}                 & 40.9\%                         & 30.7\%                         & 35.1\%                          & \multicolumn{1}{r|}{0.17}        & 37.4\%                         & 56.7\%                         & 45.1\%                          & \multicolumn{1}{r|}{0.24}        & 34.5\%                         & 85.8\%                         & 49.2\%                          & 0.20                             & 42.7\%                         & 32.5\%                         & 36.9\%                          & \multicolumn{1}{r|}{0.19}        & 39.4\%                         & 59.6\%                         & 47.4\%                          & \multicolumn{1}{r|}{\cellcolor[HTML]{FCCD7D}0.28} & 35.5\%                         & 87.2\%                         & 50.5\%                          & \cellcolor[HTML]{FCCD7D}0.22     \\ \cline{1-1}
\textbf{PolyCoder}                & 43.3\%                         & 33.8\%                         & 38.0\%                          & \multicolumn{1}{r|}{0.20}        & 38.3\%                         & 59.9\%                         & 46.7\%                          & \multicolumn{1}{r|}{0.26}        & 34.4\%                         & 86.2\%                         & 49.1\%                          & 0.20                             & 38.1\%                         & 30.1\%                         & 33.6\%                          & \multicolumn{1}{r|}{0.14}        & 36.4\%                         & 57.3\%                         & 44.5\%                          & \multicolumn{1}{r|}{0.23}                         & 35.2\%                         & 87.4\%                         & 50.2\%                          & 0.22                             \\ \hline
                                  & \multicolumn{12}{c|}{\textbf{Saliency\_reading}}                                                                                                                                                                                                                                                                                                                                                                   & \multicolumn{12}{c|}{\textbf{Saliency\_coding}}                                                                                                                                                                                                                                                                                                                                                                                     \\ \cline{2-25} 
                                  & \multicolumn{4}{c|}{\textbf{Top 5}}                                                                                                  & \multicolumn{4}{c|}{\textbf{Top 10}}                                                                                                 & \multicolumn{4}{c|}{\textbf{Top 20}}                                                                                                 & \multicolumn{4}{c|}{\textbf{Top 5}}                                                                                                  & \multicolumn{4}{c|}{\textbf{Top 10}}                                                                                                                  & \multicolumn{4}{c|}{\textbf{Top 20}}                                                                                                 \\ \cline{2-25} 
\multirow{-3}{*}{\textbf{Method}} & \multicolumn{1}{c}{\textbf{P}} & \multicolumn{1}{c}{\textbf{R}} & \multicolumn{1}{c}{\textbf{F1}} & \multicolumn{1}{c|}{\textbf{KA}} & \multicolumn{1}{c}{\textbf{P}} & \multicolumn{1}{c}{\textbf{R}} & \multicolumn{1}{c}{\textbf{F1}} & \multicolumn{1}{c|}{\textbf{KA}} & \multicolumn{1}{c}{\textbf{P}} & \multicolumn{1}{c}{\textbf{R}} & \multicolumn{1}{c}{\textbf{F1}} & \multicolumn{1}{c|}{\textbf{KA}} & \multicolumn{1}{c}{\textbf{P}} & \multicolumn{1}{c}{\textbf{R}} & \multicolumn{1}{c}{\textbf{F1}} & \multicolumn{1}{c|}{\textbf{KA}} & \multicolumn{1}{c}{\textbf{P}} & \multicolumn{1}{c}{\textbf{R}} & \multicolumn{1}{c}{\textbf{F1}} & \multicolumn{1}{c|}{\textbf{KA}}                  & \multicolumn{1}{c}{\textbf{P}} & \multicolumn{1}{c}{\textbf{R}} & \multicolumn{1}{c}{\textbf{F1}} & \multicolumn{1}{c|}{\textbf{KA}} \\ \hline
\textbf{Incoder}                  & 36.6\%                         & 29.3\%                         & 32.6\%                          & \multicolumn{1}{r|}{0.06}        & 41.9\%                         & 67.3\%                         & 51.6\%                          & \multicolumn{1}{r|}{0.29}        & 41.4\%                         & 88.6\%                         & 56.4\%                          & 0.30                             & 39.4\%                         & 31.5\%                         & 35.0\%                          & \multicolumn{1}{r|}{0.09}        & 42.8\%                         & 67.3\%                         & 52.3\%                          & \multicolumn{1}{r|}{0.29}                         & 41.3\%                         & 88.5\%                         & 56.4\%                          & 0.30                             \\ \cline{1-1}
\textbf{CodeGen}                  & 46.1\%                         & 34.9\%                         & 39.8\%                          & \multicolumn{1}{r|}{0.23}        & 39.3\%                         & 59.6\%                         & 47.4\%                          & \multicolumn{1}{r|}{0.28}        & 35.3\%                         & 87.4\%                         & 50.3\%                          & 0.23                             & 47.3\%                         & 35.4\%                         & 40.5\%                          & \multicolumn{1}{r|}{0.24}        & 40.7\%                         & 60.8\%                         & 48.8\%                          & \multicolumn{1}{r|}{0.30}                         & 36.2\%                         & 88.5\%                         & 51.4\%                          & 0.25                             \\ \cline{1-1}
\textbf{CodeParrot}               & 53.8\%                         & 42.1\%                         & 47.2\%                          & \multicolumn{1}{r|}{0.31}        & 44.6\%                         & 69.3\%                         & 54.3\%                          & \multicolumn{1}{r|}{0.38}        & 34.8\%                         & 86.8\%                         & 49.7\%                          & 0.21                             & 56.3\%                         & 43.7\%                         & 49.2\%                          & \multicolumn{1}{r|}{0.35}        & 45.3\%                         & 69.6\%                         & 54.9\%                          & \multicolumn{1}{r|}{0.39}                         & 35.7\%                         & 87.9\%                         & 50.8\%                          & 0.23                             \\ \cline{1-1}
\textbf{GPT-J-6B}                 & 41.2\%                         & 30.6\%                         & 35.1\%                          & \multicolumn{1}{r|}{0.17}        & 35.5\%                         & 53.1\%                         & 42.6\%                          & \multicolumn{1}{r|}{0.20}        & 34.2\%                         & 85.1\%                         & 48.8\%                          & 0.19                             & 41.2\%                         & 30.5\%                         & 35.0\%                          & \multicolumn{1}{r|}{0.17}        & 37.5\%                         & 56.3\%                         & 45.0\%                          & \multicolumn{1}{r|}{0.24}                         & 35.5\%                         & 87.2\%                         & 50.5\%                          & 0.22                             \\ \cline{1-1}
\textbf{PolyCoder}                & 41.6\%                         & 32.5\%                         & 36.5\%                          & \multicolumn{1}{r|}{0.18}        & 36.5\%                         & 56.9\%                         & 44.5\%                          & \multicolumn{1}{r|}{0.23}        & 34.2\%                         & 85.9\%                         & 49.0\%                          & 0.20                             & 36.7\%                         & 29.0\%                         & 32.4\%                          & \multicolumn{1}{r|}{0.13}        & 35.3\%                         & 55.6\%                         & 43.2\%                          & \multicolumn{1}{r|}{0.21}                         & 35.0\%                         & 87.0\%                         & 50.0\%                          & 0.22                             \\ \hline
\end{tabular}

}
    \end{subtable}
        \vspace{1mm}
    \end{minipage}
    % \hfill
    \begin{minipage}[t]{\linewidth}
    \begin{subtable}[h]{\textwidth}
    \setlength{\tabcolsep}{1.5pt}
    \caption{Self-attention-based methods.}
    \label{table:attention_based}
    \scriptsize
    \resizebox{\linewidth}{!}{
        \begin{tabular}{|l|rrrrrrrrrrrr|rrrrrrrrrrrr|}
\hline
                                  & \multicolumn{12}{c|}{\textbf{READING\_first}}                                                                                                                                                                                                                                                                                                                                                                                                        & \multicolumn{12}{c|}{\textbf{CODING\_first}}                                                                                                                                                                                                                                                                                                                                                                                        \\ \cline{2-25} 
                                  & \multicolumn{4}{c|}{\textbf{Top 5}}                                                                                                                   & \multicolumn{4}{c|}{\textbf{Top 10}}                                                                                                                  & \multicolumn{4}{c|}{\textbf{Top 20}}                                                                                                 & \multicolumn{4}{c|}{\textbf{Top 5}}                                                                                                  & \multicolumn{4}{c|}{\textbf{Top 10}}                                                                                                                  & \multicolumn{4}{c|}{\textbf{Top 20}}                                                                                                 \\ \cline{2-25} 
\multirow{-3}{*}{\textbf{Method}} & \multicolumn{1}{c}{\textbf{P}} & \multicolumn{1}{c}{\textbf{R}} & \multicolumn{1}{c}{\textbf{F1}} & \multicolumn{1}{c|}{\textbf{KA}}                  & \multicolumn{1}{c}{\textbf{P}} & \multicolumn{1}{c}{\textbf{R}} & \multicolumn{1}{c}{\textbf{F1}} & \multicolumn{1}{c|}{\textbf{KA}}                  & \multicolumn{1}{c}{\textbf{P}} & \multicolumn{1}{c}{\textbf{R}} & \multicolumn{1}{c}{\textbf{F1}} & \multicolumn{1}{c|}{\textbf{KA}} & \multicolumn{1}{c}{\textbf{P}} & \multicolumn{1}{c}{\textbf{R}} & \multicolumn{1}{c}{\textbf{F1}} & \multicolumn{1}{c|}{\textbf{KA}} & \multicolumn{1}{c}{\textbf{P}} & \multicolumn{1}{c}{\textbf{R}} & \multicolumn{1}{c}{\textbf{F1}} & \multicolumn{1}{c|}{\textbf{KA}}                  & \multicolumn{1}{c}{\textbf{P}} & \multicolumn{1}{c}{\textbf{R}} & \multicolumn{1}{c}{\textbf{F1}} & \multicolumn{1}{c|}{\textbf{KA}} \\ \hline
\textbf{Incoder}                  & 44.9\%                         & 38.5\%                         & 41.5\%                          & \multicolumn{1}{r|}{0.17}                         & \cellcolor[HTML]{FCCD7D}46.7\% & \cellcolor[HTML]{FCCD7D}75.9\% & \cellcolor[HTML]{FCCD7D}57.8\%  & \multicolumn{1}{r|}{\cellcolor[HTML]{FCCD7D}0.39} & \cellcolor[HTML]{FCCD7D}41.7\% & 90.2\%                         & \cellcolor[HTML]{FCCD7D}57.0\%  & \cellcolor[HTML]{FCCD7D}0.31     & 45.3\%                         & 38.0\%                         & 41.4\%                          & \multicolumn{1}{r|}{0.18}        & 45.1\%                         & 73.0\%                         & 55.7\%                          & \multicolumn{1}{r|}{0.35}                         & 41.4\%                         & 89.5\%                         & 56.7\%                          & 0.30                             \\ \cline{1-1}
\textbf{CodeGen}                  & 9.8\%                          & 8.3\%                          & 9.0\%                           & \multicolumn{1}{r|}{-0.17}                        & 26.0\%                         & 41.6\%                         & 32.0\%                          & \multicolumn{1}{r|}{0.05}                         & 34.0\%                         & 84.1\%                         & 48.4\%                          & 0.19                             & 9.5\%                          & 8.5\%                          & 9.0\%                           & \multicolumn{1}{r|}{-0.17}       & 25.6\%                         & 40.4\%                         & 31.3\%                          & \multicolumn{1}{r|}{0.04}                         & 33.8\%                         & 83.8\%                         & 48.2\%                          & 0.19                             \\ \cline{1-1}
\textbf{CodeParrot}               & 33.8\%                         & 27.5\%                         & 30.3\%                          & \multicolumn{1}{r|}{0.10}                         & 37.9\%                         & 59.9\%                         & 46.4\%                          & \multicolumn{1}{r|}{0.26}                         & 34.5\%                         & 86.2\%                         & 49.3\%                          & 0.20                             & 44.0\%                         & 35.7\%                         & 39.4\%                          & \multicolumn{1}{r|}{0.22}        & 42.2\%                         & 67.2\%                         & 51.9\%                          & \multicolumn{1}{r|}{0.35}                         & 34.6\%                         & 86.6\%                         & 49.4\%                          & 0.21                             \\ \cline{1-1}
\textbf{GPT-J-6B}                 & 6.5\%                          & 5.5\%                          & 6.0\%                           & \multicolumn{1}{r|}{-0.21}                        & 23.0\%                         & 33.2\%                         & 27.2\%                          & \multicolumn{1}{r|}{-0.03}                        & 33.0\%                         & 81.3\%                         & 46.9\%                          & 0.16                             & 7.4\%                          & 6.2\%                          & 6.7\%                           & \multicolumn{1}{r|}{-0.20}       & 22.2\%                         & 33.0\%                         & 26.5\%                          & \multicolumn{1}{r|}{-0.04}                        & 33.2\%                         & 81.9\%                         & 47.3\%                          & 0.17                             \\ \cline{1-1}
\textbf{PolyCoder}                & 41.4\%                         & 32.5\%                         & 36.4\%                          & \multicolumn{1}{r|}{0.19}                         & 41.7\%                         & 63.7\%                         & 50.4\%                          & \multicolumn{1}{r|}{0.33}                         & 35.8\%                         & 87.9\%                         & 50.8\%                          & \cellcolor[HTML]{FCCD7D}0.23     & 43.8\%                         & 33.8\%                         & 38.1\%                          & \multicolumn{1}{r|}{0.20}        & \cellcolor[HTML]{FCCD7D}42.5\% & 64.8\%                         & \cellcolor[HTML]{FCCD7D}51.3\%  & \multicolumn{1}{r|}{\cellcolor[HTML]{FCCD7D}0.34} & 35.8\%                         & 87.8\%                         & 50.9\%                          & 0.23                             \\ \hline
                                  & \multicolumn{12}{c|}{\textbf{READING\_last}}                                                                                                                                                                                                                                                                                                                                                                                                         & \multicolumn{12}{c|}{\textbf{CODING\_last}}                                                                                                                                                                                                                                                                                                                                                                                         \\ \cline{2-25} 
                                  & \multicolumn{4}{c|}{\textbf{Top 5}}                                                                                                                   & \multicolumn{4}{c|}{\textbf{Top 10}}                                                                                                                  & \multicolumn{4}{c|}{\textbf{Top 20}}                                                                                                 & \multicolumn{4}{c|}{\textbf{Top 5}}                                                                                                  & \multicolumn{4}{c|}{\textbf{Top 10}}                                                                                                                  & \multicolumn{4}{c|}{\textbf{Top 20}}                                                                                                 \\ \cline{2-25} 
\multirow{-3}{*}{\textbf{Method}} & \multicolumn{1}{c}{\textbf{P}} & \multicolumn{1}{c}{\textbf{R}} & \multicolumn{1}{c}{\textbf{F1}} & \multicolumn{1}{c|}{\textbf{KA}}                  & \multicolumn{1}{c}{\textbf{P}} & \multicolumn{1}{c}{\textbf{R}} & \multicolumn{1}{c}{\textbf{F1}} & \multicolumn{1}{c|}{\textbf{KA}}                  & \multicolumn{1}{c}{\textbf{P}} & \multicolumn{1}{c}{\textbf{R}} & \multicolumn{1}{c}{\textbf{F1}} & \multicolumn{1}{c|}{\textbf{KA}} & \multicolumn{1}{c}{\textbf{P}} & \multicolumn{1}{c}{\textbf{R}} & \multicolumn{1}{c}{\textbf{F1}} & \multicolumn{1}{c|}{\textbf{KA}} & \multicolumn{1}{c}{\textbf{P}} & \multicolumn{1}{c}{\textbf{R}} & \multicolumn{1}{c}{\textbf{F1}} & \multicolumn{1}{c|}{\textbf{KA}}                  & \multicolumn{1}{c}{\textbf{P}} & \multicolumn{1}{c}{\textbf{R}} & \multicolumn{1}{c}{\textbf{F1}} & \multicolumn{1}{c|}{\textbf{KA}} \\ \hline
\textbf{Incoder}                  & 40.2\%                         & 31.9\%                         & 35.6\%                          & \multicolumn{1}{r|}{0.12}                         & 41.2\%                         & 64.8\%                         & 50.4\%                          & \multicolumn{1}{r|}{0.27}                         & 41.0\%                         & 88.8\%                         & 56.1\%                          & 0.29                             & 41.9\%                         & 33.9\%                         & 37.5\%                          & \multicolumn{1}{r|}{0.13}        & 43.2\%                         & 69.6\%                         & 53.3\%                          & \multicolumn{1}{r|}{0.31}                         & 41.4\%                         & 89.3\%                         & 56.6\%                          & 0.30                             \\ \cline{1-1}
\textbf{CodeGen}                  & 33.8\%                         & 24.8\%                         & 28.6\%                          & \multicolumn{1}{r|}{0.08}                         & 39.5\%                         & 59.1\%                         & 47.4\%                          & \multicolumn{1}{r|}{0.29}                         & 34.7\%                         & 85.2\%                         & 49.3\%                          & 0.21                             & 31.0\%                         & 24.4\%                         & 27.3\%                          & \multicolumn{1}{r|}{0.07}        & 37.4\%                         & 57.1\%                         & 45.2\%                          & \multicolumn{1}{r|}{0.25}                         & 35.1\%                         & 86.4\%                         & 49.9\%                          & 0.22                             \\ \cline{1-1}
\textbf{CodeParrot}               & \cellcolor[HTML]{FCCD7D}60.1\% & \cellcolor[HTML]{FCCD7D}45.4\% & \cellcolor[HTML]{FCCD7D}51.7\%  & \multicolumn{1}{r|}{\cellcolor[HTML]{FCCD7D}0.38} & \cellcolor[HTML]{FCCD7D}50.4\% & \cellcolor[HTML]{FCCD7D}76.4\% & \cellcolor[HTML]{FCCD7D}60.8\%  & \multicolumn{1}{r|}{\cellcolor[HTML]{FCCD7D}0.48} & \cellcolor[HTML]{FCCD7D}37.0\% & 90.2\%                         & \cellcolor[HTML]{FCCD7D}52.4\%  & \cellcolor[HTML]{FCCD7D}0.26     & 56.4\%                         & 43.2\%                         & 48.9\%                          & \multicolumn{1}{r|}{0.35}        & 48.5\%                         & 73.6\%                         & 58.4\%                          & \multicolumn{1}{r|}{0.45}                         & 36.7\%                         & 89.7\%                         & 52.1\%                          & 0.26                             \\ \cline{1-1}
\textbf{GPT-J-6B}                 & 8.8\%                          & 7.0\%                          & 7.8\%                           & \multicolumn{1}{r|}{-0.19}                        & 26.6\%                         & 42.7\%                         & 32.8\%                          & \multicolumn{1}{r|}{0.05}                         & 33.4\%                         & 82.7\%                         & 47.5\%                          & 0.17                             & 13.8\%                         & 10.6\%                         & 12.0\%                          & \multicolumn{1}{r|}{-0.14}       & 27.4\%                         & 41.8\%                         & 33.1\%                          & \multicolumn{1}{r|}{0.06}                         & 33.6\%                         & 83.2\%                         & 47.9\%                          & 0.18                             \\ \cline{1-1}
\textbf{PolyCoder}                & 23.1\%                         & 18.1\%                         & 20.3\%                          & \multicolumn{1}{r|}{-0.03}                        & 34.3\%                         & 53.8\%                         & 41.9\%                          & \multicolumn{1}{r|}{0.19}                         & 34.7\%                         & 86.4\%                         & 49.5\%                          & 0.21                             & 32.8\%                         & 25.7\%                         & 28.9\%                          & \multicolumn{1}{r|}{0.09}        & 37.9\%                         & 58.8\%                         & 46.1\%                          & \multicolumn{1}{r|}{0.26}                         & 35.4\%                         & 87.0\%                         & 50.3\%                          & 0.22                             \\ \hline
                                  & \multicolumn{12}{c|}{\textbf{READING\_all}}                                                                                                                                                                                                                                                                                                                                                                                                          & \multicolumn{12}{c|}{\textbf{CODING\_all}}                                                                                                                                                                                                                                                                                                                                                                                          \\ \cline{2-25} 
                                  & \multicolumn{4}{c|}{\textbf{Top 5}}                                                                                                                   & \multicolumn{4}{c|}{\textbf{Top 10}}                                                                                                                  & \multicolumn{4}{c|}{\textbf{Top 20}}                                                                                                 & \multicolumn{4}{c|}{\textbf{Top 5}}                                                                                                  & \multicolumn{4}{c|}{\textbf{Top 10}}                                                                                                                  & \multicolumn{4}{c|}{\textbf{Top 20}}                                                                                                 \\ \cline{2-25} 
\multirow{-3}{*}{\textbf{Method}} & \multicolumn{1}{c}{\textbf{P}} & \multicolumn{1}{c}{\textbf{R}} & \multicolumn{1}{c}{\textbf{F1}} & \multicolumn{1}{c|}{\textbf{KA}}                  & \multicolumn{1}{c}{\textbf{P}} & \multicolumn{1}{c}{\textbf{R}} & \multicolumn{1}{c}{\textbf{F1}} & \multicolumn{1}{c|}{\textbf{KA}}                  & \multicolumn{1}{c}{\textbf{P}} & \multicolumn{1}{c}{\textbf{R}} & \multicolumn{1}{c}{\textbf{F1}} & \multicolumn{1}{c|}{\textbf{KA}} & \multicolumn{1}{c}{\textbf{P}} & \multicolumn{1}{c}{\textbf{R}} & \multicolumn{1}{c}{\textbf{F1}} & \multicolumn{1}{c|}{\textbf{KA}} & \multicolumn{1}{c}{\textbf{P}} & \multicolumn{1}{c}{\textbf{R}} & \multicolumn{1}{c}{\textbf{F1}} & \multicolumn{1}{c|}{\textbf{KA}}                  & \multicolumn{1}{c}{\textbf{P}} & \multicolumn{1}{c}{\textbf{R}} & \multicolumn{1}{c}{\textbf{F1}} & \multicolumn{1}{c|}{\textbf{KA}} \\ \hline
\textbf{Incoder}                  & 34.8\%                         & 32.5\%                         & 33.6\%                          & \multicolumn{1}{r|}{0.08}                         & 39.0\%                         & 66.7\%                         & 49.2\%                          & \multicolumn{1}{r|}{0.25}                         & 40.4\%                         & 88.2\%                         & 55.4\%                          & 0.28                             & 40.6\%                         & 39.2\%                         & 39.9\%                          & \multicolumn{1}{r|}{0.16}        & 42.0\%                         & 71.8\%                         & 53.0\%                          & \multicolumn{1}{r|}{0.31}                         & 41.2\%                         & 89.8\%                         & 56.5\%                          & 0.30                             \\ \cline{1-1}
\textbf{CodeGen}                  & 19.6\%                         & 14.7\%                         & 16.8\%                          & \multicolumn{1}{r|}{-0.09}                        & 24.9\%                         & 37.0\%                         & 29.7\%                          & \multicolumn{1}{r|}{-0.00}                        & 34.0\%                         & 83.9\%                         & 48.4\%                          & 0.19                             & 28.4\%                         & 23.1\%                         & 25.4\%                          & \multicolumn{1}{r|}{0.03}        & 32.1\%                         & 50.7\%                         & 39.3\%                          & \multicolumn{1}{r|}{0.15}                         & 34.5\%                         & 85.8\%                         & 49.2\%                          & 0.21                             \\ \cline{1-1}
\textbf{CodeParrot}               & 41.8\%                         & 31.8\%                         & 36.1\%                          & \multicolumn{1}{r|}{0.16}                         & 39.6\%                         & 61.3\%                         & 48.1\%                          & \multicolumn{1}{r|}{0.28}                         & 35.1\%                         & 87.1\%                         & 50.1\%                          & 0.22                             & 50.1\%                         & 39.7\%                         & 44.3\%                          & \multicolumn{1}{r|}{0.28}        & 43.4\%                         & 68.0\%                         & 53.0\%                          & \multicolumn{1}{r|}{0.36}                         & 35.2\%                         & 87.3\%                         & 50.1\%                          & 0.22                             \\ \cline{1-1}
\textbf{GPT-J-6B}                 & 12.6\%                         & 10.2\%                         & 11.3\%                          & \multicolumn{1}{r|}{-0.15}                        & 22.9\%                         & 35.0\%                         & 27.7\%                          & \multicolumn{1}{r|}{-0.04}                        & 33.8\%                         & 83.6\%                         & 48.1\%                          & 0.18                             & 30.6\%                         & 24.4\%                         & 27.2\%                          & \multicolumn{1}{r|}{0.05}        & 32.7\%                         & 51.2\%                         & 39.9\%                          & \multicolumn{1}{r|}{0.15}                         & 34.6\%                         & 85.6\%                         & 49.2\%                          & 0.21                             \\ \cline{1-1}
\textbf{PolyCoder}                & 28.7\%                         & 21.3\%                         & 24.4\%                          & \multicolumn{1}{r|}{0.01}                         & 36.3\%                         & 56.3\%                         & 44.1\%                          & \multicolumn{1}{r|}{0.21}                         & 34.8\%                         & 86.5\%                         & 49.6\%                          & 0.21                             & 33.4\%                         & 25.7\%                         & 29.0\%                          & \multicolumn{1}{r|}{0.08}        & 38.4\%                         & 59.7\%                         & 46.8\%                          & \multicolumn{1}{r|}{0.26}                         & 35.1\%                         & 86.9\%                         & 50.0\%                          & 0.22                             \\ \hline
\end{tabular}
    }

    \end{subtable}
    \end{minipage}
\end{table*}

The order of different attention calculation methods is randomized to mitigate the learning effect.
We also do not reveal the names of the attention calculation methods to reduce bias.
At the end of the user study, participants answer three open-ended questions about different attention calculation methods and the user study design. 
We ask these open-ended questions to study the correlation between model explainability and user trust. These questions include:

\begin{enumerate}[label={(\arabic*)}]
    \item[Q4] \textit{Are you interested to know how the LLM generates the code?}
    \item[Q5] \textit{What do you want to find out about the internal code generation process in LLM?}
    \item[Q6] \textit{Do you trust this LLM? What do you need to know to improve the trust of the LLM?}
\end{enumerate}

% ------------------------------------------------------

% -----------------------RESULTS------------------------
\section{RESULTS}
\subsection{RQ1: To What Extent Is Model Attention Aligned With Human Attention?} 
\label{sec:5.1}
To answer this question, we collect the top $K$ keywords that the six LLMs attend to and compare them with the keywords labeled as important in the programmer attention dataset. As described in Section~\ref{sec:attention_alignment}, we use \mr{\sm{}~\cite{da2019fine} and \ka{}~\cite{krippendorff2018content}} to measure the attention alignment between LLMs and human programmers. When running models where the max output length can be set, we set the token limit to the number of tokens of the ground truth plus 20 to tolerate moderate redundancy.

Table~\ref{table:pertubation_based}, Table \ref{table:gradient_based}, and Table \ref{table:attention_based} present the attention alignment results when computing attention scores with \textit{perturbation-based}, \textit{gradient-based}, and \textit{self-attention-based} methods, respectively. Because GPT-4 does not reveal its internal states during runtime, only perturbation-based methods are applicable. Therefore, this section discusses the results of the best perturbation-based method, \textit{BERT\_masking}.
Section~\ref{sec:rq3} compares different attention calculation methods in detail.

With \textit{BERT\_masking} method, from $K=5$ to $K=10$, the F1 scores of all models increase because as more and more tokens are selected by the models, we observe very high (around 90\%) recall, which compensates for the declining precision. However, from $K=10$ to $K=20$, the F1 scores remain stable due to rapid declines in precision scores. On the other hand, the \ka{} scores remain stable from $K=5$ to $K=10$ (except for GPT-4) but rapidly decrease from $K=10$ to $K=20$. 

Overall, for all models and all $K$ values, the \ka{} does not change much and remains below 0.3, and the F1 scores remain below 0.6, indicating little agreement between model attention and human attention~\cite{mchugh2012interrater}.
Among these models, GPT-4 showed the lowest attention overlap with human attention regarding both metrics. One possible explanation is that ultra-large models such as GPT-4 have developed a reasoning strategy different from that of human programmers. 
% Another possible reason is that the perturbation-based methods become less effective since GPT-4 is more spontaneous than other models in this study. For example, GPT-4 may generate different outputs for different inputs, not because the difference is significant but because of its unpredictable nature.
These results suggest a consistent attention misalignment between LLMs and programmers when generating code. 

\begin{finding}{} {\textit{There is a consistent misalignment between LLM attention and programmer attention in all settings, indicating that LLMs do not reason programming tasks like human programmers.}
}\end{finding}

\subsection{{RQ2: Can Attention Explain Errors of Code Generation Models?}}
\label{sec:errors}
To answer this question, the first two authors manually analyzed code generation errors \mr{made by the best two models in our study---GPT-4 and CodeGen-2.7B. In total, these two models generated 920 incorrect code solutions on the two benchmarks. We randomly sampled 211  incorrect solutions, including 172 incorrect solutions from CodeGen-2.7B and 39 incorrect solutions from GPT-4. The sample size is statistically significant with a 90\% confidence level and 5\% margin of error.} 
%\mr{Because only the perturbation-based methods apply to GPT-4, the manual analysis is performed with \textit{BERT\_Masking}}.

\mr{The first two authors started with the first 50 code generation errors and independently checked whether the attention pattern of each code calculated by the \textit{BERT\_masking}, which gives the most aligned results in the quantitative experiments, could explain the error in it. For simple tasks, it takes about five minutes to check each one. For complicated tasks (e.g., tasks that require an understanding of specific math concepts), it takes around 10 to 15 minutes, since the authors need to manually debug the code and inspect its runtime values to understand the error first. After analyzing 50 errors, they
discussed these errors with the other authors and summarized six common attention patterns that can be used to explain code generation errors:}

\begin{itemize}
    \item \mr{\textit{Missing attention to critical conditions.} The task description mentions certain conditions or corner cases to handle. However, the model misses or incorrectly handles one or more such conditions since it does not attend to the words or phrases that describe the corresponding conditions.}
    \item \mr{\textit{Missing attention to important descriptive words of an operation or a data object.} The task description mentions an important property of an operation or a data object, such as ``{\em largest} element'' and ``{\em ascending} order''. However, the model does not attend to these descriptive words and thus generates a code solution with incorrect logic.}
    \item \mr{\textit{Missing attention to operation descriptions.} The task description mentions an operation or action, such as {\em open a file} and {\em sort a list}. However, the model fails to attend to the verb words or phrases. Therefore, the generated code performs the wrong action or does not perform the action.}
    \item \mr{\textit{Missing attention to data types.} Programming tasks often explicitly mention the expected type of inputs and outputs. Some task descriptions also mention the data type of some intermediate results. However, the model does not attend to some data type descriptions. This can lead to different types of errors, e.g., returning the wrong type of data, calling a method on the wrong type of object, etc.}
    % \item \mr{\textit{Recurring attention}: The model starts to repeat itself after some point. This type of error usually occurs due to the limited attention window.}
    \item \mr{\textit{Incorrect mapping between NL words and code elements.} In some cases, we observe the model attends to an important word or phrase correctly but the model maps it to a wrong method call, variable, parameter, value, or logic, potentially due to some conceptual misunderstanding of the semantics meaning of the NL words.}
\end{itemize}

\mr{Then, they independently labeled the remaining errors, discussed their labeling with each other, and resolved the conflicts. %The average agreement level between their labels is 0.83 before discussion, as measured in Cohen’s Kappa score.
In total, we found that errors in 57 of the 211 incorrect solutions (27\%) can be explained by one of the five attention patterns mentioned above. Specifically, 54 of the 172 incorrect solutions from CodeGen-2.7B (31\%) are explainable and 3 of the 39 incorrect solutions from GPT-4 (8\%) are explainable. This finding suggests that neural attention analysis can be potentially applied to locate and repair a non-trivial portion of errors in LLM-generated code. Given that only 3 errors made by GPT-4 can be explained by attention misalignment patterns, this implies that weaker models such as CodeGen-2.7B are more likely to be affected by attention misalignment and thus their generation errors are more explainable. This is an interesting observation since GPT-4 also suffers from attention misalignment, as shown in Table~\ref{table:overall_result}, but its generation errors cannot be easily explained by attention analysis. This indicates that as the language models become large enough, they may have developed a different way of interpreting input prompts and generating content. This calls for new research to understand why ultra-large language models such as GPT-4 generate incorrect code.  
Besides, we acknowledge that many errors cannot be easily explained by model attention. Such errors include syntax errors, undefined variable names, incorrect API usage, incorrect array index, infinite loop, etc. This result indicates that further analysis is required to understand the root causes of these errors.} 

We show the distribution and examples for each type of error below. In these examples, we highlight keywords with high attention scores from the model in \hlblue{blue}.

\mr{\subsubsection{\textit{Missing Attention to Critical Conditions.}} 
This prompt asks for a function that matches a string with a character \textit{``a''} followed by zero or more character \textit{``b''}s. However, CodeGen-2.7B generates a function that only matches words with one \textit{``a''} followed by one or more \textit{``b''}s. Our attention analysis shows conditional phrase \textit{``followed by zero or more''} do not contain any of the most attended tokens, which explains why the model ignores the condition where the character \textit{``a''} can be followed with no \textit{``b''}s or more than one \textit{``b''}. 6 incorrect solutions fall into this category.}
\begin{lstlisting}[language=Python, escapechar=!, numberstyle=\tiny\color{lightgray}]
# !\modelkeyword{32.3}{Write}! a function that !\modelkeyword{32.3}{matches}! !\modelkeyword{32.3}{a}! !\modelkeyword{32.3}{string}! that has an ``a'' followed by zero or more ``b''!\modelkeyword{32.3}{s}!.
def function(string):
    return string.count('ab') >= 1
\end{lstlisting}

\mr{\subsubsection{Missing Attention to Important Descriptive Words of an Operation or a Data Object.} The prompt below asks for a function that calculates the $n^{th}$ power of all elements in a list. However, the code generated by CodeGen-2.7B calculates the $n^{th}$ power of a specific element without using the lambda function. Our attention analysis shows the property phrase \textit{``individual items''} contains none of the most attended words in the prompt. This explains why the model fails to perform the requested operation on all items instead of a specific item.} 13 incorrect solutions fall into this category.

\begin{lstlisting}[language=Python, escapechar=!, numberstyle=\tiny\color{lightgray}]
# Write !\bothkeyword{34.5}{{a}}! function to find the !\bothkeyword{34.5}{{n-th}}! !\bothkeyword{34.5}{{power}}! of individual elements in !\bothkeyword{34.5}{{a}}! !\bothkeyword{34.5}{{list}}! using lambda function.
def function(list, n):
    return list[n-1] ** n
\end{lstlisting}

% \mr{\subsubsection{Missing attention to important descriptive words of an operation or a data object.} The example below asks for a function that returns elements that appear singly in each tuple from a tuple list. The code generated by GPT-4 misunderstands this prompt to return elements that only appear once in the entire tuple list. Our attention analysis shows GPT-4 fails to attend to the keyword \textit{``singly''}, which partly explains this error. 13 errors belong to this category.}
% \begin{lstlisting}[language=Python, escapechar=!, numberstyle=\tiny\color{lightgray}]
% # Write !\modelkeyword{32.3}{a}! !\modelkeyword{32.3}{function}! to !\modelkeyword{32.3}{extract}! elements that occur singly in !\modelkeyword{32.3}{the}! given !\modelkeyword{32.3}{tuple}! list.
% def extract_singly(tuple_list):
%     elements = [item for sublist in tuple_list for item in sublist]
%     counts the occurrences of each element
%     occurrence_count = {}
%     for element in elements:
%         if element in occurrence_count:
%             occurrence_count[element] += 1
%         else:
%             occurrence_count[element] = 1
%     result = [element for element, count in occurrence_count.items() if count == 1]
%     return result
% \end{lstlisting}

\mr{\subsubsection{Missing Attention to Operation Descriptions.} The prompt below asks for a function that counts how many pairs of elements at the same index in two given arrays are equal. However, CodeGen-2.7B generates a function that checks whether the two strings are equal. Our attention analysis shows the model the operator keyword \textit{``count''} is not one of the most attended words in the prompt, which explains why the generated code does not perform counting as requested. 6 incorrect solutions fall into this category.}

\begin{lstlisting}[language=Python, escapechar=!, numberstyle=\tiny\color{lightgray}]
# Write !\modelkeyword{32.3}{a}! function to count the same !\modelkeyword{32.3}{pair}! !\modelkeyword{32.3}{in}! two given !\modelkeyword{32.3}{lists}! using !\modelkeyword{32.3}{map}! function.
def function(x,y):
    return x == y
\end{lstlisting}

% \begin{lstlisting}[language=Python, escapechar=!, numberstyle=\tiny\color{lightgray}]
% # !\modelkeyword{32.3}{Write}! a !\modelkeyword{32.3}{function}! to replace !\modelkeyword{32.3}{whitespaces}! with an underscore and vice !\modelkeyword{32.3}{versa}! in a given string by using !\modelkeyword{32.3}{regex}!.
% def replace_spaces(input_string):
%     string_with_underscores = re.sub(r'\s', '_', input_string)
%     result_string = re.sub(r'_', ' ', string_with_underscores)
%     return result_string
% \end{lstlisting}

\mr{\subsubsection{Missing Attention to Data Types.} This prompt asks for a function that matches a word that contains \textit{``z''} but does not have \textit{``z''} at the start or end. However, CodeGen-2.7B generates a function that checks whether the string retrieved from removing the first character from the given word equals the string retrieved from removing the last character from the given word. Our attention analysis shows that the data type keyword \textit{``z''} is not one of the most attended words in the prompt. This explains why the generated code fails to check for the existence of a specific character in the given word. 7 errors belong to this category.}

\begin{lstlisting}[language=Python, escapechar=!, numberstyle=\tiny\color{lightgray}]
# !\modelkeyword{32.3}{Write}! a !\modelkeyword{32.3}{function}! that !\modelkeyword{32.3}{matches}! !\modelkeyword{32.3}{a}! word !\modelkeyword{32.3}{containing}! 'z', not at the start or end of the word.
def function(word):
    return word[1:] == word[:-1]
\end{lstlisting}

% \begin{lstlisting}[language=Python, escapechar=!, numberstyle=\tiny\color{lightgray}]
% # Write !\modelkeyword{32.3}{a}! !\modelkeyword{32.3}{function}! !\modelkeyword{32.3}{to}! check if the given tuple contains all !\modelkeyword{32.3}{valid}! !\modelkeyword{32.3}{values}! or not.
% def check_valid(tup):
%     return all(isinstance(item, int) and item is not None for item in tup)
% \end{lstlisting}

\mr{\subsubsection{Incorrect Mapping Between NL Words and Code Elements.} This prompt asks for a function that calculates the $n^{th}$ bell number. However, instead of invoking the correct formula for bell number, CodeGen-2.7B calculates the factorial of $n$. Our attention analysis shows the keyword \textit{``bell''} is one of the most attended words in the prompt. However, it still generates incorrectly for this prompt because it does not understand the semantic meaning of \textit{``bell number''}. 25 incorrect solutions belong to this category.}
\begin{lstlisting}[language=Python, escapechar=!, numberstyle=\tiny\color{lightgray}]
# !\modelkeyword{32.3}{Write}! a !\modelkeyword{32.3}{python}! function to !\modelkeyword{32.3}{find}! nth !\modelkeyword{32.3}{bell}! !\modelkeyword{32.3}{number}!.
def function(n):
    if n == 1:
        return 1
    else:
        return n * function(n - 1)
\end{lstlisting}

\begin{finding}{} {
\textit{Attention analysis can explain \mr{27}\% of generation errors in a benchmark of \mr{211} errors. These errors provide insights into the limitations of LLMs.}
}\end{finding}

\subsection{{RQ3:} What Is the Impact of Attention Calculation Methods on the Alignment?} 

% \begin{table*}[t]
%     \caption{The average San Martino's F-1 score and Krippendorff's Alpha over all five models (excluding GPT-4) in various $K$ settings. The highest score in each column is highlighted in \yellow{yellow}.}
%     \label{table:comparison}

%     \resizebox{0.5\linewidth}{!}{
%         \input{table/comparison}
%     }
% \end{table*}

\label{sec:rq3}

To compare different attention calculation methods, we calculated the average San Martino's F-1 score and \ka{} of all five models (except for GPT-4) for all $K$ (Table~\ref{table:comparison}).
Table~\ref{table:comparison} shows that \textit{BERT\_masking} gives the highest alignment in both metrics for $K=5,10$. However, the other perturbation-based method, \textit{SHAP}, does not outperform other methods as \textit{BERT\_masking} did. 
On the other hand, gradient-based methods are generally better than self-attention-based methods, especially for $K=5$ in terms of both metrics. Surprisingly, attention scores computed by self-attention-based methods (e.g., \textit{CODING\_first}) are least aligned with human attention, especially for smaller $K$ values. One possible reason is that other computation units in the transformer architecture, such as the feed-forward layers, also play an important role in the code generation process. For instance, Geva et al.~find that feed-forward layers influence model predictions by promoting concepts in the vocabulary space~\cite{geva2022transformer}. Thus, only considering self-attention layers does not fully capture the influence of each input token on model predictions. 

\begin{wraptable}{r}{0.48\linewidth}  % 'r' for right, 'l' for left, and specify width
    \caption{The average San Martino's F-1 score and Krippendorff's Alpha over all five models (excluding GPT-4) in various $K$ settings. The highest score in each column is highlighted in \yellow{yellow}.}
    \label{table:comparison}
    \resizebox{\linewidth}{!}{
        \begin{tabular}{|l|rr|rr|rr|}
\hline
                                        & \multicolumn{2}{c|}{\textbf{Top 5}}                                & \multicolumn{2}{c|}{\textbf{Top 10}}                               & \multicolumn{2}{c|}{\textbf{Top 20}}                               \\ \cline{2-7} 
\multirow{-2}{*}{\textbf{Method}}       & \multicolumn{1}{c}{\textbf{F1}} & \multicolumn{1}{c|}{\textbf{KA}} & \multicolumn{1}{c}{\textbf{F1}} & \multicolumn{1}{c|}{\textbf{KA}} & \multicolumn{1}{c}{\textbf{F1}} & \multicolumn{1}{c|}{\textbf{KA}} \\ \hline
\textbf{BERT\_masking}             & \cellcolor[HTML]{FCCD7D}45.7\%  & \cellcolor[HTML]{FCCD7D}0.27     & \cellcolor[HTML]{FCCD7D}51.4\%  & 0.26                             & \cellcolor[HTML]{FCCD7D}52.1\%  & 0.12                             \\
\textbf{SHAP}                           & 30.6\%                          & 0.11                             & 41.5\%                          & 0.18                             & 48.0\%                          & 0.16                             \\ \hline
\textbf{Input$\times$Gradient\_reading} & 37.9\%                          & 0.21                             & 47.3\%                          & 0.30                             & 48.2\%                          & 0.22                             \\
\textbf{Input$\times$Gradient\_coding}  & 38.3\%                          & 0.22                             & 47.9\%                          & \cellcolor[HTML]{FCCD7D}0.31     & 49.2\%                          & \cellcolor[HTML]{FCCD7D}0.24     \\
\textbf{Saliency\_reading}              & 37.5\%                          & 0.20                             & 46.0\%                          & 0.28                             & 48.0\%                          & 0.22                             \\
\textbf{Saliency\_coding}               & 37.5\%                          & 0.21                             & 46.7\%                          & 0.29                             & 49.0\%                          & \cellcolor[HTML]{FCCD7D}0.24     \\ \hline
\textbf{READING\_first}                 & 24.6\%                          & 0.01                             & 42.7\%                          & 0.20                             & 50.5\%                          & 0.22                             \\
\textbf{READING\_last}                  & 28.8\%                          & 0.07                             & 46.6\%                          & 0.26                             & 51.0\%                          & 0.23                             \\
\textbf{READING\_all}                   & 24.4\%                          & 0.00                             & 39.8\%                          & 0.14                             & 50.3\%                          & 0.22                             \\
\textbf{CODING\_first}                  & 26.9\%                          & 0.05                             & 43.4\%                          & 0.21                             & 50.5\%                          & 0.22                             \\
\textbf{CODING\_last}                   & 30.9\%                          & 0.10                             & 47.2\%                          & 0.27                             & 51.4\%                          & \cellcolor[HTML]{FCCD7D}0.24     \\
\textbf{CODING\_all}                    & 33.2\%                          & 0.12                             & 46.4\%                          & 0.25                             & 51.0\%                          & 0.23                             \\ \hline
\end{tabular}  % Make sure this file contains a table
    }
\end{wraptable}
Furthermore, within gradient-based methods, we observe that calculating attention distribution during the coding stage (e.g., \textit{Saliency\_coding}) gives better alignment than the attention distribution during the reading stage (e.g., \textit{Saliency\_reading}). This pattern can be observed for all $K$ values (i.e., $K=5,10$). 
However, the differences between \textit{Input$\times$Gradient} and \textit{Saliency} are trivial.

Finally, as discussed in Section~\ref{sec:self_attention_based_methods}, there is no consensus on how to aggregate self-attention scores~\cite{zhang2022does, zeng2022extensive, bensemann2022eye}. We experimented with six different self-attention aggregation methods used in previous work to find the best self-attention aggregation method. First, among the three options---(1) only summing attention scores in the first layer, (2) only summing attention scores in the last layer, and (3) summing attention scores from all layers---only summing attention scores in the last layer produces the attention distributions that are most aligned with human attention distributions in both reading and coding stages, except when K is set to 5 in the coding stage. This suggests that future research should consider the last layer when leveraging self-attentions to interpret code language models. 
Similar to the finding in gradient-based methods, for self-attention-based methods, attention distributions at the coding stage (e.g., \textit{CODING\_first}) are consistently more aligned with human attention than those at the reading stage (e.g., \textit{READING\_first}).

\begin{finding}{} {\textit{\textit{BERT\_masking} produces the best attention alignment to human programmers among all methods.  \textit{Gradient-based methods} are generally better than self-attention-based methods. For gradient-based and self-attention-based methods, attention distributions in the coding stage produce higher alignment. For self-attention-based methods, attention distributions in the last layer produce the highest alignment.}
}\end{finding}

\subsection{{RQ4:} Which Attention Calculation Method Is the Most Preferred?} 

%\begin{wrapfigure}{r}{0.55\textwidth} % "r" for right side, "0.55\textwidth" for width
\begin{figure}[t]
  \centering
  \includegraphics[width=0.7\linewidth]{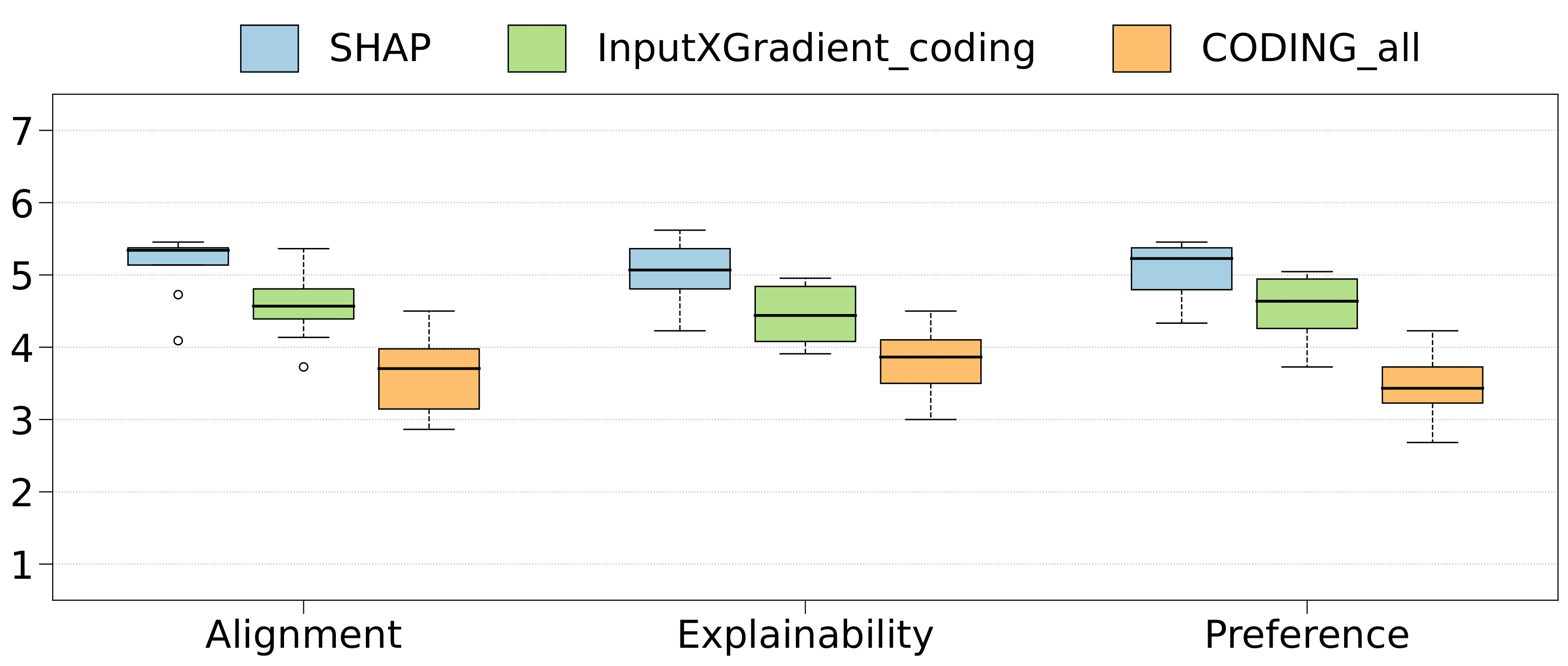}
  \caption{Participants' choices over different attention calculation methods in three dimensions.}
  \label{fig:user_response}
\end{figure}
%\end{wrapfigure}
Figure~\ref{fig:user_response} shows participants' assessment about \textit{SHAP}, \textit{Input$\times$Gradient\_coding}, and \textit{CODING\_all} in terms of the alignment with their own attention, the explainability of the computed attention, and their own preference (Q1-Q3 in Section~\ref{sec:user-study-design}). Overall, the perturbation-based method, \textit{SHAP}, is favored over the other two methods in all three aspects. The average rating on the attention alignment of \textit{SHAP} is 5.13, while the average for \textit{InputXGradient\_coding} and \textit{CODING\_all} are 4.59 and 3.62, respectively. 
% The mean differences between \textit{SHAP} and \textit{InputXGradient\_coding} and between \textit{SHAP} and \textit{CODING\_all} are both statistically significant (Welch's $t$-test, $p=0.05$ and $p=0.004$). 

Furthermore, participants suggested that \textit{SHAP} has better explainability than \textit{InputXGradient\_coding} (mean difference: 0.59, Welch's $t$-test, $p=0.02$) and \textit{CODING\_all} (mean difference: 1.26, Welch's $t$-test, $p=0.00001$). However, according to participants' responses about their trust in LLMs (Q6 in Section~\ref{sec:user-study-design}), 14 out of 22 participants expressed a lack of confidence and trust, even after seeing the attention-based explanations. For example, P3 wrote, \textit{``I want to know whether the LLM model can provide the reasons why they generate the code. For instance, the reference code.''} Participants also asked for more fine-grained \mr{attention analysis}. Specifically, they wished to see which parts of the input are responsible for generating which parts of the output. For instance, P10 said, \textit{``[I want to know] how the LLM determines the input and output, what are the discriminative words that drive different generations.''} Participants also showed more preference for \textit{SHAP} over the other two methods. The mean preference differences between \textit{SHAP} and \textit{InputXGradient\_coding} and between \textit{SHAP} and \textit{CODING\_all} is 5.04 vs.~4.56 (Welch's $t$-test, $p=0.05$) and 5.04 vs.~3.48 (Welch's $t$-test, $p=0.00009$).

\begin{finding}{} {\textit{Overall, participants preferred the perturbation-based method over the gradient-based and the self-attention-based methods. However, participants still felt a lack of trust in LLMs after seeing the attention-based explanations and wished to see richer explanations such as reference code and fine-grained attention mapping between text and code.}
}\end{finding}
% ------------------------------------------------------

% --------------------IMPLICATIONS----------------------
\section{IMPLICATIONS AND OPPORTUNITIES}
Our study has several significant implications that benefit the development of more reliable and more accurate LLMs for code generation in the future.

\mr{In RQ1, we discovered a consistent misalignment between human and model attention in all six models. While it is possible that LLMs use an entirely different method to reason about task descriptions compared with human programmers, it is arguable whether such a method is indeed good for programmers due to concerns about interpretability, robustness, and trust. Many studies have shown that human-aligned models are perceived as more trustworthy by humans~\cite{huang2021attributes, gao2022aligning, bansal2023towards, stocco2022thirdeye, hazard2022importance, kotseruba2016joint}. Furthermore, in practice, LLMs today can reliably solve a limited number of simple programming tasks and struggle to handle more sophisticated or custom tasks, which indicates a huge space for improvement. Thus, we believe investigating the attention patterns of LLMs is a worthwhile effort to help us understand how LLMs generate code and why LLMs make some mistakes and also inform new opportunities to improve LLMs.}

%\mr{With our results,} future researchers \original{to} \mr{can} understand \original{model predictions} \mr{model attention patterns} better and leverage them for more accurate and robust code generation. %Current prompt engineering approaches are mostly model-agnostic, which means they do not consider the unique characteristics of code generation models~\cite{liu2023pre}. For example, a typical way to craft an effective prompt is to decompose the task in the prompt into different steps and make the instructions as specific as possible. But should there be differences in how to craft prompts for general LLMs and LLMs for code generation? What accommodations should be made for different code generation models? With our study, future researchers may analyze patterns in keywords attended to by different models. Then, they can develop better prompt engineering approaches for each model by substituting keywords in the original prompts with words that are more likely to be attended by the model.

\mr{In RQ2, we manually analyzed the attention of 211 incorrect generations of the best two models in our paper (GPT-4 and CodeGen-2.7B) and found 27\% of the errors can be explained by incorrect attention alignment.} Our finding suggests the potential of fixing these generation errors by adjusting the attention alignment. Similarly, previous work in Computer Vision has shown that the performance of neural models can be improved if we force their attention to align with humans~\cite{fong2018using, jia2018biometric, melicio2018object, nunes2020learning}. \mr{One potential solution is to extend the loss function of LLMs with a penalty term that measures the KL divergence between model attention and human attention. During training, the loss will increase when model attention deviates from human attention. As a result, the LLM will be trained to distribute attention in a way similar to human programmers.} We have open-sourced our human attention dataset to facilitate future work on attention alignment. 

\mr{Furthermore, the model attention patterns in RQ3} can help to improve the robustness of code generation models by developing new attention-based adversarial training methods. Most adversarial training methods only apply small perturbations to random tokens in a prompt, without considering the importance of a token to the model~\cite{yefet2020adversarial, zhang2020generating, bielik2020adversarial, srikant2021generating}. Our results show that it is worthwhile to prioritize important tokens during perturbation. Perturbing these important words and asking the model to generate code for these perturbed prompts may reveal robustness issues more efficiently.

\mr{In RQ3, we compared 12 attention calculation methods with quantitative experiments (Table~\ref{table:overall_result}).} \mr{Therefore}, our study provides practical guidelines for choosing the attention calculation method for future research on LLM-based code generation models. Our experiment results indicate developers who want to measure the attention of code generation models should first consider \textit{BERT\_masking} since it consistently achieves the best alignment with human attention in all except for two settings (Table~\ref{table:comparison}). \mr{Between self-attention-based and gradient-based methods}, researchers should prioritize gradient-based methods that generally give better and more stable alignment in most settings (Table~\ref{table:comparison}). For gradient-based and self-attention-based methods, calculating the attention distribution of the coding stage gives better alignment (Table~\ref{table:comparison}). Finally, when leveraging self-attention to interpret transformer-based code generation models, researchers should consider using the self-attention scores computed from the last layer, which is demonstrated to be more aligned with human attention (Table~\ref{table:comparison}).    
% However, for self-attention-based methods, calculating attention distribution at the reading stage gives better alignment (Table~\ref{table:comparison}). 

\mr{In RQ4, we conducted a user study to compare developers' perceptions on different explanation methods. Most participants considered SHAP as the best XAI method for LLM code generation models. This finding suggests that future researchers may want to use the perturbation-based method as the default XAI method for LLM-based code generation. Furthermore, in the post-study survey, participants also asked for more fine-grained attention analysis, such as revealing the association between individual input and output tokens. This highlights the need for new XAI methods to interpret LLM-based code generation models.}

\mr{In the future, we would also like to explore how to teach humans how LLMs interpret code. Understanding how LLMs interpret code can inspire human programmers to craft prompts that are more understandable by LLMs and thus guide LLMs to generate better code.}
% ------------------------------------------------------

% ----------------THREATS TO VALIDITY-------------------
\section{THREAT TO VALIDITY}
\noindent\textbf{Internal validity.}
{One potential threat lies in the manual labeling process. Two programmers manually labeled the important words to understand a task description and implement the correct function. 
Since this criterion of keywords is highly subjective, the words selected by the two labelers may not represent the choice of a large pool of programmers. To mitigate this threat, the authors established a labeling standard through discussion and achieved substantial agreement on the labeling. Furthermore, we invite a third labeler to validate our annotated dataset by independently labeling the 164 prompts from the HumanEval dataset. We calculated the Fleiss' Kappa score among the labels of three labelers and the result (0.64) indicates a significant agreement.}

\noindent\textbf{External validity.} One potential threat to external validity is that we have only experimented with one programming language, Python. We cannot guarantee that our findings generalize to another language. 

\noindent\textbf{Construct validity.} 
{One potential threat to construct validity lies in the survey design. As we know, programming problems are mentally demanding to solve and programmers may have different views on which part of the prompt should the model attend to. However, in the current design, only 22 participants were involved. The small sample size may result in findings that are not generalizable. To combat this threat, we only invited participants who are experienced in Python programming and have used code generation LLMs at least once to control the quality of the user study.}
% ------------------------------------------------------

% -------------------RELATED WORK-----------------------
\section{RELATED WORK}
\subsection{Code Generation From Natural Language}
{Since CodeBERT~\cite{feng2017component}, there has been a large body of literature where Large Language Models (LLMs) are used in code generation~\cite{guo2019towards, feng2020codebert, zeng2020recparser, shen2022incorporating,  wang2020rat, chen2021codex, wang2021codet5, gpt-neox-library, parvez2021retrieval, tunstall2022natural, fried2022incoder, xu2022systematic, nijkamp2022codegen}. Both Guo et al.~\cite{guo2019towards} and Zeng et al.~\cite{zeng2020recparser} used a pre-trained BERT~\cite{devlin2018bert} model to encode NL questions and database schemas for text-to-SQL generation. CodeBERT adopts the same model architecture as BERT but is trained with a hybrid objective on code and text data from GitHub repositories~\cite{feng2017component}. Codex, CodeGPT, and GraphCodeBERT improve CodeBERT by leveraging data flow in the pre-training stage~\cite{guo2020graphcodebert}. Recently, modern LLMs such as GPT-4 and Google Bard excel in code generation tasks on various benchmarks~\cite{bubeck2023sparks, destefanis2023preliminary}. The highly accurate and contextually rich code snippets generated by these models enable the automation of various programming tasks.}

In addition to developing new LLMs for code, prior work has also presented new methods to enhance LLMs for more accurate and robust code generation~\cite{parvez2021retrieval, zhang2022diet, shen2022incorporating, chen2022codet, chakraborty2022natgen, zan2022cert, li2023skcoder}. Instead of directly generating code from text, REDCODER~\cite{parvez2021retrieval} first retrieves similar code snippets from a corpus and then passes them along with the text description to an LLM for code generation. Shen et al.~\cite{shen2022incorporating} proposed to leverage domain knowledge from documentation and code comments to assist code generation. Chakraborty et al.~proposed a new pre-training task called {\em naturalizing of source code} (i.e., translate an artificially created code to a human-written form) to help an LLM learn how to generate natural code~\cite{chakraborty2022natgen}. Chen et al.~proposed to use an LLM to automatically generate test cases to examine the code generated by the same LLM~\cite{chen2022codet}. Zan et al.~proposed to first decompose a LLM to two LLMs, one for generating program sketches and the other for filling the sketches~\cite{zan2022cert}.  SkCoder~\cite{li2023skcoder} is designed to mimic developers' code reuse behavior by first retrieving a code example related to a given task, extracting a sketch from it, and editing the sketch based on the task description.

\subsection{Empirical Studies on Code Generation}
Recently, many studies have evaluated LLM-based code generation models in different aspects, including performance~\cite{liu2020empirical, chen2021codex, hendrycksapps2021, rodrigues2021studying,  zeng2022extensive}, robustness~\cite{zhuo2023robustness, liu2023reliability}, security~\cite{asare2022github, pearce2022asleep, yetistiren2022assessing}, code smells~\cite{siddiq2022empirical}, usability~\cite{vaithilingam2022expectation, xu2022systematic, barke2022grounded, bird2022taking}, and licensing~\cite{ciniselli2022extent}. 
The most related to us are those that investigate the explainability of LLMs for code~\cite{karmakar2021pre, liguori2022can, zhang2022does, wan2022they}. Karmakar et al.~studied what pre-trained BERT-based models such as CodeBERT and GraphCodeBERT have learned about code using probing tasks~\cite{karmakar2021pre}. A probing task is essentially a prediction task on a specific code property, such as code length and cyclomatic complexity, to test whether the model is sensitive to the property. Compared with our work, they did not analyze the code generation process, e.g., why and how certain code is generated based on a text description. Liguori et al.~proposed a perturbation-based method to evaluate NMT models for code generation~\cite{liguori2022can}.
In addition, Zhang et al.~investigated the code generation process by analyzing the self-attention layers in CodeBERT and GraphCodeBERT~\cite{zhang2022does}. Wan et al.~did a similar self-attention analysis and also designed a new probe task on code structures to analyze whether LLMs have learned information about code structures~\cite{wan2022they}.
Compared with these studies, our work differs by analyzing the consistency between LLMs' attention and programmers' attention on the NL description of a programming task.

\subsection{Model Attention Analysis in Other Domains}
\mr{Many attention calculation methods have been developed to explain models in other domains. For example, in CV domain, Selvaraju et al.~\cite{selvaraju2017grad} proposed to use the gradients of a convolutional neural network (CNN) to indicate the importance of each pixel in an image for a specific prediction. 
% This technique was widely adopted and improved by subsequent works~\cite{draelos2020use, chattopadhay2018grad, wang2020score, ramaswamy2020ablation}.
Similarly, Zhou et al.~\cite{zhou2016learning} used the global average pooling mechanism to calculate the importance of each region for CNN to classify an image correctly. Autonomous driving is another domain where the need for explainability is strong. For example, Zeiler et al.~\cite{zeiler2014visualizing} use deconvolution layers to
understand how autonomous vehicles capture real-time image segments using CNNs. In another work, Chen et al.~\cite{chen2021interpretable} proposed a
data-efficient policy learning approach called Semantic Predictive Control (SPC) that explains how perceived environmental states are mapped to actions.}
% ------------------------------------------------------

% ---------------------CONCLUSION-----------------------
\section{CONCLUSION}
This paper presents an empirical study on attention alignment between LLM-based code generation models and human programmers. Our results reveal that there is a consistent misalignment between LLMs' and programmers' attention. Among the 12 attention calculation methods, perturbation-based methods produced attention scores that were better aligned with human attention and were also more preferred by user study participants. Based on our study results, we further discuss several implications and future research opportunities for better interpretation and performance improvement of LLM-based code generation models.
% ------------------------------------------------------

% -----------------DATA AVAILABILITY--------------------
\section{DATA AVAILABILITY}
Our code and data are available on a GitHub repository~\cite{attention_alignment_empirical_study}.
% ------------------------------------------------------

\bibliographystyle{ACM-Reference-Format}
\bibliography{reference}
\end{document}